\documentclass[prx,singlecolumn,english,superscriptaddress,floatfix,longbibliography,nofootinbib, aps, 11pt]{revtex4-2}

\usepackage{amsmath, amssymb,bbm, bm, braket, dsfont, times, amsthm}
\usepackage{graphicx,xfrac}
\usepackage[margin=1in]{geometry}
\usepackage[dvipsnames]{xcolor}
\usepackage[breaklinks,colorlinks,allcolors=frenchrose]{hyperref}
\hypersetup{
    colorlinks=true,
}
\usepackage{framed}
\usepackage{cleveref}

\usepackage{tikz-cd}

\usetikzlibrary{positioning,intersections}
\usetikzlibrary{calc}

\usepackage[normalem]{ulem}
\usepackage{mdframed}
\usepackage{setspace}
\usepackage{array}

\renewcommand{\>}{\rangle}
\newcommand{\<}{\langle}

\newcommand{\rep}{\mathbf{Rep}}
\newcommand{\To}{\Rightarrow}

\newcommand{\Mod}{\mathbf{Mod}}
\newcommand{\Bimod}{\mathbf{Bimod}}
\renewcommand{\hom}{\mathbf{Hom}}
\newcommand{\vc}{\mathbf{Vec}}

\newcommand{\id}{\mathbf{id}}

\newcommand{\alg}{\mathbf{Alg}}

\newcommand{\Aut}{\text{Aut}}
\newcommand{\symto}{$\text{SymTFT}$~}
\newcommand{\phys}{\mathbf{phys}}
\newcommand{\sym}{\mathbf{sym}}
\newcommand{\IR}{\mathbf{IR}}

\usepackage{enumitem}
\usepackage{euscript}
\newcommand{\e}[1]{\EuScript{#1}}
\newcommand{\fk}[1]{\mathfrak{#1}}

\newcommand{\bbm}[1]{\mathbbm{#1}}
\newcommand{\cc}[1]{\mathcal{#1}}

\newcommand{\Z}{\mathcal{Z}}
\newcommand{\D}{\cc{D}}

\newcommand{\bZ}{\mathbb{Z}}
\newcommand{\bC}{\mathbb{C}}

\newcommand{\edo}{\text{End}}
\newcommand{\imm}{\text{im}}
\newcommand{\cg}[2]{{\vphantom{#2}}^{#1}#2} 
\newcolumntype{C}[1]{>{\centering\let\newline\\\arraybackslash\hspace{0pt}}m{#1}}

\newcommand{\hide}[1]{}

\newtheoremstyle{mystyle}
  {3pt}
  {3pt}
  {\it}
  {}
  {\bfseries}
  {:}
  {.5em}
  {}
\theoremstyle{mystyle}
\newtheorem{theorem}{Theorem}[section]
\newtheorem{ptheorem}{Theorem$^\mathbf{ph}$}[section]
\newtheorem{definition}{Definition}[section]
\newtheorem{lemma}{Lemma}[section]
\newtheorem{proposition}{Proposition}[section]
\newtheorem{corollary}{Corollary}[section]

\newtheorem*{fact}{Fact}
\theoremstyle{remark}
\newtheorem*{remark}{Remark}
\newtheorem{example}{Example}[section]

\definecolor{zima}{HTML}{5BC2E7}
\definecolor{bleudefrance}{rgb}{0.19, 0.55, 0.91}
\definecolor{cyan(process)}{rgb}{0.0, 0.72, 0.92}
\definecolor{jasper}{rgb}{0.84, 0.23, 0.24}
\definecolor{frenchrose}{rgb}{0.96, 0.29, 0.54}
\definecolor{folly}{rgb}{1.0, 0.0, 0.31}
\definecolor{junglegreen}{rgb}{0.16, 0.67, 0.53}
\definecolor{mediumseagreen}{rgb}{0.24, 0.7, 0.44}
\definecolor{blue1}{HTML}{116278}
\definecolor{blue2}{HTML}{7CA8C3}
\definecolor{pink1}{HTML}{e6c9cc}
\definecolor{red1}{HTML}{f1a3a5}
\definecolor{yellow1}{HTML}{edd5b0}
\definecolor{orange1}{HTML}{edae84}
\makeatletter
\def\l@subsubsection#1#2{} 
\makeatother
\begin{document}
\title{\Large Topological Holography for 2+1-D Gapped and Gapless Phases with Generalized Symmetries}
\author{Rui Wen} 
\affiliation{Department of Physics and Astronomy, and Stewart Blusson Quantum Matter Institute, University of British Columbia, Vancouver, BC, Canada V6T 1Z1}
\date{\today}   
\begin{abstract}
We study topological holography for 2+1-D gapped and gapless phases with generalized symmetries using tools from higher linear algebra and higher condensation theory. We focus on bosonic fusion 2-category symmetries, where the  Symmetry Topological Field Theory (\symto) are 3+1D Dijkgraaf-Witten theories.

(1). Gapped phases are obtained from the sandwich construction with gapped symmetry and physical boundaries. A gapped boundary of the 3+1D \symto is called minimal if it has no intrinsic 2+1-D topological order. We derive the general structure of a sandwich construction with minimal gapped symmetry and physical boundaries, including the underlying topological order and the symmetry action. We also study some concrete examples with 2-group or non-invertible symmetries. 

(2). For gapless phases, we show that the \symto provides a complete description of the \textit{topological skeleton} of a gapless phase. The topological skeleton of a gapless phase is the higher categorical structure of its topological defects. We rigorously establish this relation for 2+1-D gapless phases with finite group symmetries. For a gapless phase with a finite group symmetry, its topological skeleton(also known as gapless SPT(gSPT)) can be characterized by the decorated domain wall construction. We give a precise formulation of this using spectral sequence. We show that certain class of condensable algebras in the \symto $\Z_1[2\vc_G]$, which we call minimal condensable algebras, has exactly the same structure. Our result naturally categorifies the previously known relation in 1+1D. We further give a cohomological classification of minimal condensable algebras, which enables us to compute the classification of 2+1-D $G$-gSPTs via ordinary group cohomology. Finally we use \symto to construct 2+1-D gSPT with generalized symmetries, including an  intrinsically gSPT(igSPT) with exact non-invertible fusion 2-category symmetry and anomalous 2-group IR symmetry.
\end{abstract}
\maketitle
\newpage
\tableofcontents
\section{Introduction}
The interplay between topology, symmetry and phases of matter is one of the most fascinating subjects of study in condensed matter theory and mathematical physics. To date, a reasonably complete theory for gapped phases of matter has been established~\cite{Kong_2020,Johnson_Freyd_2022,gaiotto2019condensationshighercategories,Lan_2017,Lan_2018,Lan_2019,kong2014braidedfusioncategoriesgravitational,kong2015boundarybulkrelationtopologicalorders,Kong_2022,lurie2009classificationtopologicalfieldtheories,Wen_2023}. Up to invertible phases, a gapped phase of matter is described by the multi-fusion higher category of its topological defects.~\footnote{Here we only consider quantum liquids, excluding non-liquid phases such as fractons. } Symmetry can interact with gapped phases, giving rise to the notion of symmetry-enriched topological phases~\cite{Barkeshli_2019,Lan_2016,Barkeshli_2022,Ye_2024}. Moreover, the very concept of symmetry has been reshaped over the past decade. While traditionally described by a group acting globally, it is now understood that a symmetry can be any topological operator within a theory—one that may not act on the entire system and might not even be invertible~\cite{Heidenreich:2021xpr,
Kaidi:2021xfk, Choi:2021kmx, Roumpedakis:2022aik,Bhardwaj:2022yxj, Choi:2022zal,Cordova:2022ieu,Choi:2022jqy,Kaidi:2022uux,Antinucci:2022eat,Bashmakov:2022jtl,Damia:2022bcd,Choi:2022rfe,Bhardwaj:2022lsg,GarciaEtxebarria:2022vzq,Heckman:2022muc,Niro:2022ctq,Kaidi:2022cpf,Antinucci:2022vyk,Chen:2022cyw,Lin:2022dhv,Bashmakov:2022uek,Karasik_2023,cordova2022neutrinomassesgeneralizedsymmetry,Chang_2023,etxebarria2022goldstonetheoremcontinuousnoninvertible,D_coppet_2023,moradi2022topologicalholographyunificationlandau,Runkel_2023,choi2023noninvertiblegausslawaxions,Bhardwaj_2023,Bhardwaj_2023_2,Bartsch_2024,heckman2023top,antinucci2024zoologynoninvertibledualitydefects,apte2023obstructions,delcamp2024higher,kaidi2023symmetry,li2023noninvertible,brennan2024coupling,etheredge2023branessymmetriesmathcaln3,lin2023bootstrapping,putrov2024categorical,carta2023comments,koide2023noninvertible,zhang2023anomalies,cao2023subsystem,dierigl2024r7,inamura2024fusion,chen2023symtfts,bashmakov2023four,choi2023remarks,bhardwaj2023generalized,pace2023exact,van2023monopoles,lawrie2024intermediate,apruzzi2024aspects,chen2023solitonic,cordova2024quantum,sun2023duality,pace2024emergent,cordova2024anomalies,antinucci2023anomalies,duan2023zn,chen2023fusion,nagoya2023non,cordova2024axion,copetti2025higher,damia2024non,Bhardwaj_2025,bhardwaj2024fusion3categoriesdualitydefects,bhardwaj2025latticemodelsphasestransitions}. See ~\cite{schafernameki2023ictplecturesnoninvertiblegeneralized,shao2024whatsundonetasilectures,Luo_2024,brennan2023introductiongeneralizedglobalsymmetries} for reviews on generalized symmetries.  Such generalized symmetries are naturally described by fusion higher-categories.  The study of gapped phases with generalized symmetries is therefore intimately related to the higher representation theory of higher fusion categories.
\begin{figure}[h]
    \centering
        \begin{tikzpicture}[baseline=(current  bounding  box.center)]
\draw[thick]  (0,0) -- (0,4);
\draw[thick]  (2,0) -- (2,4);
\fill[zima] (0,0) rectangle (2,4);
\node[label, below] at (0,0){$\fk{B}_\phys$};
\node[label,left] at (0,2){$\e{D}$};
\node[label, below] at (2.2,0){$\fk{B}_\sym$};
\node[label, right] at (2,2){$\e{C}$};
\node[label] at (1,2){$\Z_1[\e{C}]$};
\node[label] at (3,2){$\simeq$};
\node[label] at (4.2,1.94){$\e{D}\boxtimes_{\Z_1[\e{C}]}\e{C}$};
\end{tikzpicture}
    \caption{The sandwich construction,  mathematically given by a relative tensor product.}
    \label{fig: sandwich_1}
\end{figure}

For studying gapped phases with invertible symmetries, one can often avoid the higher category theory language by turning to homotopy or (co-)homology theory. On the other hand, it is generally a challenging task to construct and analyze properties of models with non-invertible symmetries. The symmetry topological field theory(\symto) provides a systematic method for this task~\cite{chatterjee2024emergentgeneralizedsymmetrymaximal,Kong_2020,Kong_2020_ahc,Kong_2022_cql1,Kong_2024_cql2,Kong_2022_1d,moradi2022topologicalholographyunificationlandau,kaidi2023symmetrytftsnoninvertibledefects,bhardwaj2024gappedphases21dnoninvertible,kaidi2023symmetrytftsanomaliesnoninvertible,bhardwaj2024boundarysymtft}.  This method is based on the so-called sandwich construction, shown in Fig.~\ref{fig: sandwich_1}, and the powerful theory of bulk-boundary relation of topological orders~\cite{kong2015boundarybulkrelationtopologicalorders,Kong_2017,Kong_2018,Kong_2020_gapless1,Kong_2021_gapless2,Kong_2022_cql1,Kong_2024_cql2}. In this construction, a $d+1$-D theory with symmetry $\e{C}$(which is generally a fusion $d$-category) is obtained via an interval compactification of a topological order in $d+2$-D. This $d+2$-D topological order,called the symmetry topological field theory, is given by the Drinfeld center $\Z_1[\e{C}]$ and lives in the bulk of the sandwich. One crucial property of the theory $\Z_1[\e{C}]$ is that it admits a canonical gapped boundary where the topological defects form exactly $\e{C}$. This is taken to be one of the two boundaries of the sandwich, called the symmetry boundary. The other boundary of the sandwich, called the physical boundary, can be arbitrary and the sandwich produces a 2+1-D phase with symmetry $\e{C}$ upon interval compactification. Therefore, to construct a gapped phase with symmetry $\e{C}$, one simply needs to construct a gapped boundary of the \symto $\Z_1[\e{C}]$. This latter task is often more under control. 

More recently it was discovered that the \symto paradigm extends beyond gapped phases and describes certain topological properties of gapless phases as well~\cite{wen2023classification11dgaplesssymmetry,huang2023topologicalholographyquantumcriticality,Chatterjee_2023,bottini2024gaplessphasehaagerupsymmetry,antinucci2025symtft31dgaplessspts,bhardwaj2024hassediagramsgaplessspt,bhardwaj2024clubsandwichgaplessphases,wen2024topologicalholographyfermions,wen2025stringcondensationtopologicalholography,bhardwaj2024fermionicnoninvertiblesymmetries11d,huang2024fermionicquantumcriticalitylens}. Even in a gapless phase, there can be topological defects and it is expected that they form certain higher category structure. In~\cite{Kong_2018,Kong_2020_gapless1,Kong_2021_gapless2,Kong_2022_cql1,Kong_2024_cql2}, this higher category structure of topological defects was named the ``topological skeleton" of a gapless phase. One standard example of this is the ``gapless symmetry protected topological phase"(gSPT)~\cite{Wen_2023,Li_2024,li2023intrinsicallypurelygaplesssptnoninvertibleduality,Scaffidi_2017,Thorngren_2021,Yu_2024}. For a symmetry group $G$, a $G$-gSPT is a gapless phase where $G$ does not act faithfully in the IR. Instead, there exists a normal subgroup $N$, and the action  of $G$ factors through $G/N$ in the IR. This means there is a gapped sector in the system such that any excitation transforming nontrivially under $N$ is necessarily in the gapped sector. For this reason $N$ is called the gapped symmetry of the gSPT, and $G/N$ is called the IR symmetry of the gSPT. One intriguing feature of gSPTs is that the IR symmetry $G/N$ can be anomalous, as long as the anomaly is trivialized in $G$. This anomaly of $G/N$ is called the emergent anomaly of the gSPT, and the gSPT is  called an intrinsically gSPT(igSPT) if the emergent anomaly is nontrivial. Categorically, we may write the structure of a ($d+1$-D)gSPT as 
\begin{align}
    d\vc_G\to d\vc_{G/N}^\omega\xrightarrow{\text{acts on}} \text{IR Theory},
\end{align}
where the first arrow is a (linear monoidal $d$-)functor that describes how the exact symmetry $G$ reduces to the IR symmetry $G/N$(with anomaly $\omega$), and the second arrow describes the faithful action of $G/N$ on the IR theory. 
The symmetry extension structure of a gSPT allows it to have nontrivial topological defects. By analyzing the structure of these topological defects, a classification of 1+1D $G$-gSPT was proposed in~\cite{wen2023classification11dgaplesssymmetry}. Somewhat surprisingly, it was found that condensable algebras in the \symto $\Z_1[\vc_G]$ have exactly the same classification. This surprising relation turns out to be part of the full \symto paradigm. 

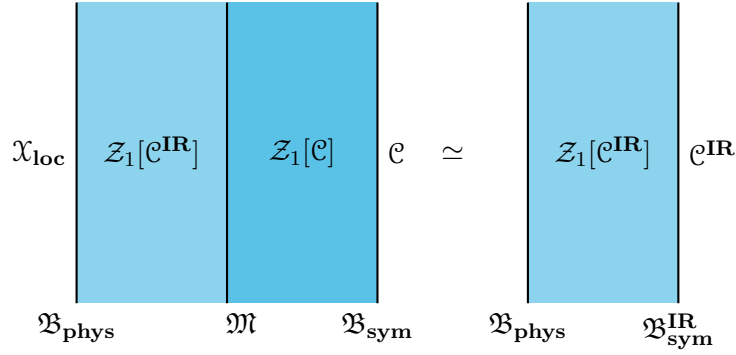
\begin{figure}
    \centering
        \begin{tikzpicture}[baseline=(current  bounding  box.center)]
\fill[zima!70] (0,0) rectangle (2,4);
\fill[zima] (2,0) rectangle (4,4);
\draw[ thick]  (0,0) -- (0,4);
\draw[ thick] (4,0) -- (4,4);
\draw[thick]  (2,0) -- (2,4);
\node[label, below] at (0,0){$\fk{B}_{\mathbf{phys}}$};
\node[label,left] at (0,2){$\e{X}_{\mathbf{loc}}$};
\node[label, below] at (2.2,0){$\fk{M}_{}$};
\node[label, below] at (4,0){$\fk{B}_\sym$};
\node[label] at (1,2){$\Z_1[\e{C}^\IR]$};
\node[label] at (3,2){$\Z_1[\e{C}]$};
\node[label, right] at (4,2){$\e{C}$};
\node[label] at (5,2){$\simeq$};
\fill[zima!70] (6,0) rectangle (8,4);
\draw[thick] (6,0) -- (6,4);
\draw[thick] (8,0) -- (8,4);
\node[label, below] at (6,0){$\fk{B}_\phys$};
\node[label, below] at (8,0){$\fk{B}^{\mathbf{IR}}_{\sym}$};
\node[label] at (7,2){$\Z_1[\e{C}^\mathbf{IR}]$};
\node[label, right] at (8,2){$\e{C}^\mathbf{IR}$};
\end{tikzpicture}
    \caption{The club-sandwich construction captures the topological skeleton of a gapless phase with symmetry. The left physical boundary encodes all the local information of the gapless phase such as OPE of local fields. The domain wall $\fk{M}_{}$ encodes the topological skeleton of the gapless phase. By fusing the domain wall $\fk{M}_{}$ with the symmetry boundary $\fk{B}_{\sym}$, we obtain an ordinary sandwich that describes the IR symmetry of the gapless phase. This process defines a functor $\e{C}\to \e{C}^{\mathbf{IR}}$.}
    \label{fig: club_sandwich_intro}
\end{figure}

In the \symto construction for gapped phases, we consider the physical boundary to be gapped, then it is a standard result of the bulk-boundary relation theory that a gapped boundary of a topological order is determined by a Lagrangian algebra~\cite{Kong_2014,Kong_2017}. On the other hand, a gapless boundary is generally associated with a non-Lagrangian condensable algebra~\cite{Chatterjee_2023,Kong_2020_gapless1,Kong_2021_gapless2,wen2023classification11dgaplesssymmetry}. This is best illustrated via the so-called ``club-sandwich" construction~\cite{bhardwaj2024clubsandwichgaplessphases,bhardwaj2024hassediagramsgaplessspt}, see Fig.~\ref{fig: club_sandwich_intro}. In this construction, we divide the bulk into two regions separated by a gapped domain wall $\fk{M}$. The right region is the \symto $\Z_1[\e{C}]$ with its symmetry boundary $\e{C}$. The left region $\Z_1[\e{C}^\mathbf{IR}]$ is obtained from the \symto through a condensation, which is determined by a condensable algebra $\e{A}$. The physical boundary, now living on the left boundary of the left region, is a gapless boundary with a faithful action by $\e{C}^\mathbf{IR}$. The club-sandwich describes a gapless phase with symmetry structure
\begin{align}
    \e{C}\to \e{C}^\mathbf{IR}\xrightarrow{\text{acts on}}\text{IR Theory},\label{eq: intro_gapless}
\end{align}
where the first arrow is determined by the condensable algebra $\e{A}$ and describes how the exact symmetry $\e{C}$ reduces to $\e{C}^{\mathbf{IR}}$ in the IR; the second arrow describes the faithful action of $\e{C}^{\mathbf{IR}}$ on the IR theory. The physical boundary is gapped precisely when $\e{C}^{\mathbf{IR}}$ is trivial, in which case $\e{A}$ is a Lagrangian algebra. This picture leads to a general principle of topological holography: a gapped $d+1$-D phase with symmetry $\e{C}$ corresponds to a Lagrangian algebra in the \symto $\Z_1[\e{C}]$, and \textit{the topological skeleton} of a gapless $d+1$-D phase with symmetry $\e{C}$ is determined by  a non-Lagrangian condensable algebra. We emphasis the notion of topological skeleton here because the \symto construction does not determine the second arrow in Eq.~\eqref{eq: intro_gapless}, only constraining it. Following this principle, 1+1D gSPTs with non-invertible and/or fermionic symmetries have been constructed in~\cite{bottini2024gaplessphasehaagerupsymmetry,bhardwaj2024hassediagramsgaplessspt,wen2024topologicalholographyfermions,bhardwaj2024fermionicnoninvertiblesymmetries11d,huang2024fermionicquantumcriticalitylens}. 

The \symto paradigm has been rigorously established for 1+1-D phases, owing to the well-developed frameworks of (braided) fusion category theory and anyon condensation theory in 2+1-D topological orders. In this work, we aim to systematically extend the understanding of topological holography  to 2+1-D gapped and gapless phases, leveraging recent advancements in fusion 2-category theory~\cite{douglas2018fusion2categoriesstatesuminvariant,Kong_2020_center,D_coppet_2023_alg,D_coppet_2023_Morita,décoppet2024drinfeldcentersmoritaequivalence,D_coppet_2023_gauging,D_coppet_2024,décoppet2024classificationfusion2categories,D_coppet_2025} and higher condensation theory~\cite{kong2025highercondensationtheory}. 
\subsection{Summary of results} 
\subsubsection{Gapped 2+1-D phases}
We study gapped 2+1-D phases with bosonic fusion 2-categorical symmetries via the sandwich construction. We formulate the interval compactification of the sandwich as a relative tensor product, which allows us to freely compute properties of the obtained 2+1-D gapped phase. A class of gapped boundaries of the \symto is called minimal gapped boundaries, which physically correspond to those gapped boundaries with no intrinsic topological order. For the case where the symmetry and the physical boundaries are both minimal gapped boundaries, we give a complete description of the obtained 2+1-D gapped phase using tools from higher linear algebra, including the underlying topological order and the symmetry action. We find that the obtained 2+1-D gapped phase generally has multiple vacuua, and each vacuua hosts a 2+1-D Dijkgraaf-Witten theory. We also study concrete examples where the \symto is $\Z_1[2\vc_{\bZ_2}]$, $\Z_1[2\vc_{S_3}]$ or $\Z_1[2\vc_{S_4}]$. The first two cases have been discussed recently in~\cite{bhardwaj2024gappedphases21dnoninvertible,bhardwaj2025gappedphases21dnoninvertible}, although through different methods. The case $\Z_1[2\vc_{S_4}]$ is previously unknown and contains some new features. In particular, we find
\begin{itemize}
    \item It admits a non-invertible minimal symmetry boundary that is neither a 2-group symmetry $2\vc_{\cc{G}}^\omega$ nor a 2-representation symmetry $2\rep(\cc{G})$. In this sense it is \textbf{intrinsically non-invertible}. 
    \item This symmetry admits a partial symmetry-broken phase with two vacuua, with one vacuum having the $S_3$-gauge theory and the other having the $\bZ_2$-gauge theory. 
    \item This symmetry admits two families of SPT phases that are not gauge-related to any single family of invertible SPTs. These SPTs are in this sense \textbf{intrinsically non-invertible SPTs.}
\end{itemize}
\subsubsection{Gapless 2+1-D phases}
To date, little research has focused on topological holography for 2+1-D gapless phases, and even in the case of invertible symmetries, a systematic understanding remains elusive. This  is largely attributable to the complexity of condensable algebras in 3+1-D topological orders. In this work we take on this challenge and establish a rigorous duality between condensable algebras in the \symto $\Z_1[2\vc_G]$ and 2+1-D $G$-gSPTs. To achieve this, we need to understand both sides of the duality systematically. 

The condensable algebras\footnote{More precisely condensable $E_2$-algebras, which describe condensation of strings.} in $\Z_1[2\vc_G]$ are described by $G$-crossed braided fusion categories. However we do not expect all of them to correspond to 2+1-D $G$-gSPTs. From the club-sandwich picture(Fig.~\ref{fig: club_sandwich_intro}), certain condensable algebra will induce (2+1-D)topological order on the domain wall $\fk{M}$. Such a club-sandwich describes a gapless 2+1-D system whose gapped sector has nontrivial topological order. This leads to the intriguing concept of ``gapless symmetry enriched topological order"(gSET). Naturally, we define a 2+1-D gSPT to be a special kind of gSET whose gapped sector has no topological order. This is completely in parallel with the gapped SET/SPT relation. This class of condensable algebras in $\Z_1[2\vc_G]$ that we expect to be dual to 2+1-D $G$-gSPT are called minimal condensable algebras~\footnote{The Lagrangian case exactly corresponds to the minimal gapped boundaries, hence the name.}.  Mathematically, such a condensable algebra is a $G$-crossed braided fusion category $\cc{B}$ whose trivial component $\cc{B}_e$ is trivial. Equivalently, we may say $\cc{B}$ is a $G$-crossed extension of $\vc$. However, the grading of $\cc{B}$ is generally not faithful but only supported on a normal subgroup $N\lhd G$. The grading of the condensable algebra is faithful precisely when the algebra is Lagrangian. The class of minimal condensable algebras supported on $N$ is then denoted as $G\mathbf{CrExt}(\vc,N)$, which stands for ``$G$-crossed extensions of $\vc$ with support $N$". This is the key subject on the \symto side: Elements of it are presumably  dual to 2+1-D $G$-gSPTs. We reveal a spectral sequence structure of minimal condensable algebras:
\begin{theorem}\label{thm: minimal_classification}
    Elements in $G\mathbf{CrExt}(\vc,N)$ are parameterized by $\omega\in \cc{H}^3[N,\bC^\times]$ such that $d_2\omega=0,d_3\omega=0$ together with a solution $(\gamma,p)$~\footnote{Each condition $d_i\omega=0$ means certain equation defined on the $E_2$-page has a solution. See Sec.~\ref{sec: LHS} for more details.}. Denote the corresponding condensable algebra as $\vc_N^{(\omega,\gamma,p)}$, the condensed phase is
     \begin{align*}
\Mod_{\vc_N^{(\omega,\gamma,p)}}^0(\Z_1[2\vc_G])\simeq \Z_1[2\vc_{G/N}^{d_4\omega}].
    \end{align*}
    
    Here $d_i$ are the differentials in the Lyndon–Hochschild–Serre(LHS) spectral sequence, and by $d_4\omega$ we mean a representative of it in $\cc{H}^4[G/N,\bC^\times]$ computed using the solution $(\gamma,p)$.
\end{theorem}

Furthermore, $G\mathbf{CrExt}(\vc,N)$ is naturally an abelian group and we prove an exact sequence relating it to ordinary group cohomology.
\begin{theorem}
    There is an exact sequence
    \begin{align*}
          \cc{H}^3[G/N,\bC^\times]\xrightarrow{p^*_3}  \cc{H}^3[G,\bC^\times]\xrightarrow{\cc{R}} G\mathbf{CrExt}(\vc,N)\xrightarrow{\cc{O}_4} \cc{H}^4[G/N,\bC^\times]\xrightarrow{p^*_4}\cc{H}^4[G,\bC^\times].
    \end{align*}
    
    Here $p_3^*,p_4^*$ are induced by projection $p: G\to G/N$.
\end{theorem}
If we accept that minimal condensable algebras are dual to $G$-gSPTs(which we will prove later), the above sequence can be written in physical terms as 
 \begin{align*}
     \mathbf{SPT}_{3D}^{G/N}\to\mathbf{SPT}^G_{3D}\to \mathbf{gSPT}^{(G,N)}_{3D}\to \mathbf{SPT}_{4D}^{G/N}\to \mathbf{SPT}_{4D}^{G}.
 \end{align*}
 
 This in fact has simple physical interpretation. Take the right half of the sequence as an example. For a $G$-gSPT, the IR symmetry $G/N$ may be anomalous, this anomaly is the image in $\mathbf{SPT}^{G/N}_{4D}$. However this anomaly is an emergent one, meaning that it trivializes when pulled back to $G$. Conversely, for every such emergent anomaly, one can construct a $G$-gSPT à la Wang-Wen-Witten~\cite{Wang_2018}. This means the right half of the sequence is exact. The physical meaning of the rest part of the sequence is also straightfoward and will be explained in Sec.~\ref{sec: minimal_cd_alg}.
\begin{table}[h]
    \centering\small
    \def\arraystretch{2}
    \begin{tabular}{|C{1.5cm}|C{1.5cm}|C{3.5cm}|C{3cm}|}
    \hline
      {$G$}  & $N$ &$\mathbf{gSPT}_{3D}^{G,N}$ & \text{igSPT?}\\
       \hline
        $\bZ_4$ & $\bZ_2$ & $\bZ_4$ &$\times$\\
        \hline
        $\bZ_{pq}$ & $\bZ_p$ & $\bZ_{p(p,q)}$&$\times$\\
        \hline
         $\bZ_2^a\times \bZ_4^b$ & $\bZ_2^b$ & $(\bZ_2\times \bZ_4)\times \bZ_2$&$\surd$\\
         \hline
          $\bZ_{n_1}^a\times\bZ_{pq}^b$ & $\bZ_p^b$ & $(\bZ_{p(p,q)}\times\bZ_{(p,n_1)})\times \bZ_{(p,,q,n_1)}$  & iff $(p,q,n_1)\neq 1$\\
          \hline
          $S_3$& $\bZ_3$& $\bZ_3$&$\times$\\
          \hline
          $D_4$& $\bZ_2$ &$\bZ_4\times \bZ_2^2$ & $\surd$\\
          \hline
    \end{tabular}
    \caption{Classification of 2+1-D gSPT with exact symmetry $G$ and gapped symmetry $N$ and whether the symmetry allows for igSPT.}
    \label{table: egs}
\end{table}

This exact sequence allows us to compute the classification $G$-gSPTs. The exactness of the sequence implies~\footnote{This is an equality as sets.}
\begin{align}
    \mathbf{gSPT}_{3D}^{G,N}=\underbrace{\cc{H}^3[G,\bC^\times]/\imm(p^*_3)}_{\text{weak gSPT}}\times \underbrace{\ker(p_4^*)}_{\text{emergent anomaly}}\label{eq: gSPT_classification}
\end{align}

The first factor comes from $\cc{H}^3[G,\bC^\times]$ and physically corresponds those gSPTs that are obtained by tuning a gapped $G$-SPT to a $G/N$-critical point. Historically, these were the earliest known examples of gSPTs~\cite{Scaffidi_2017,Li_2024} and were often referred to as weak or non-intrinsic. The reason for the quotient by $\imm(p_3^*)$ is that, if a $G$-SPT comes from a $G/N$-SPT, then as it is tuned to a $G/N$-critical point, all the topological defects from the $G/N$-SPT are lost and it can not be distinguished from a trivial gSPT. The second factor comes from $\cc{H}^4[G/N,\bC^\times]$, it captures precisely the emergent anomalies of $G$-gSPTs. 

Readers familiar with the decorated domain wall(DDW) construction of SPTs may have noticed the similarity between the spectral sequence structure of $G\mathbf{CrExt}(\vc,N)$ and the structure of 2+1-D $G$-SPT. Indeed, in the standard DDW construction for a 2+1-D $G$-SPT with $N\lhd G$ a normal subgroup, a decoration pattern is determined by a $\omega\in\cc{H}^3[N,\bC^\times]$ such that $d_2\omega=0,d_3\omega=0,d_4\omega=0$ together with a solution. However, as we argue in Sec.~\ref{sec: physics_of_gSPT}, for a 2+1D $G$-gSPT we no longer demand $d_4\omega=0$. This is because the quotient $G/N$ acts faithfully on the gapless degrees of freedom and is allowed to be anomalous. The $d_4\omega$ obstruction is in fact exactly the emergent anomaly of the gSPT. Therefore, the structure of $G\mathbf{CrExt}(\vc,N)$ matches precisely with the structure of 2+1-D $G$-gSPTs with gapped symmetry $N$. Furthermore, the twist of the condensed phase, $d_4\omega$, is exactly the emergent anomaly of the 2+1-D $G$-gSPT.  This establishes a rigorous correspondence between minimal condensable algebras in the \symto $\Z_1[2\vc_G]$ and 2+1-D $G$-gSPTs, which naturally categorifies the results in~\cite{wen2023classification11dgaplesssymmetry}. We summarize it as the following physical theorem.

\begin{ptheorem}
    A 2+1D $G$-gSPT with gapped symmetry $N$ and domain wall decoration data $\omega,\gamma,p$ corresponds to the minimal condensable algebra $\vc_N^{(\omega,\gamma,p)}$ in the \symto $\Z_1[2\vc_G]$. The emergent anomaly of the $G$-gSPT is equal to the twist of the condensed phase.
\end{ptheorem}

Using the formula Eq.~\eqref{eq: gSPT_classification}, we compute the classification of $G$-gSPTs for several $G$ and $N$. The results are summarized in Table.~\ref{table: egs}.

Finally, we exploit the \symto framework to construct 2+1D gapless phases with generalized symmetries. Using the club-sandwich construction, we construct 2+1-D igSPTs with 2-group or non-invertible symmetries, and emergent anomalous 2-group symmetry $2\vc_{\bZ_2^{[0]}\times \bZ_2^{[1]}}^\omega$.
\subsection{Structure of the article}
 Sec.~\ref{sec: fusion 2-cat review} reviews basics of fusion 2-categories and describes the fusion 2-categorical symmetries that will appear in the rest of the work.
 
 Sec.~\ref{sec: symtft gapped} through Sec.~\ref{sec: symtft gapped example} studies topological holography for gapped phases. In Sec.~\ref{sec: symtft gapped}, we discuss generalities of sandwich construction for 2+1-D gapped phases with symmetry. We give a mathematical formulation of the sandwich construction in terms of relative tensor product between $E_1$-module 2-categories.
 In Sec.~\ref{sec: symtft gapped example}, we discuss sandwich construction with minimal gapped boundaries. We give a systematic description of all sandwiches with minimal gapped symmetry and physical boundaries. We then study several examples, providing details of the obtained 2+1-D gapped phases, including the underlying TQFT and the symmetry action.
 
 Sec.~\ref{sec: gapless} through Sec.~\ref{sec: gSPT_2} studies topological holography for gapless phases.
  In Sec.~\ref{sec: gapless}, we review the concept of topological skeleton, relating it to the \symto club-sandwich construction and condensable algebras in the \symto.
In Sec.~\ref{sec: higher condensation}, we review condensation theory in 3+1D topological orders, focusing on string condensations, which are described by condensable $E_2$-algebras.
 In Sec.~\ref{sec: string cond}, we focus on condensable algebras in $\Z_1[2\vc_G]$, describing the structure of such condensable algebras and provide a systematic method for determining the condensed phase.
  In Sec.~\ref{sec: minimal_cd_alg}, we study a class of condensable algebras in $\Z_1[2\vc_G]$ that we call minimal condensable algebras. These condensable algebras are expected to be dual to 2+1-D $G$-gSPTs. We provide a cohomological description based on the LHS spectral sequence. In Sec.~\ref{sec: physics_of_gSPT}, we provide a pure 2+1-D analysis of 2+1-D $G$-gSPTs based the decorated domain wall construction and find perfect matching with the structure of minimal condensable algebras. Finally in Sec.~\ref{sec: gSPT_2}, we construct examples of 2+1-D gSPT with generalized symmetries.

\section{Bosonic fusion 2-categories and their SymTFT\label{sec: fusion 2-cat review}}
The full definition of fusion 2-category can be found in ~\cite{douglas2018fusion2categoriesstatesuminvariant}. Fusion 2-categories come in two types, called bosonic and fermionic. For any fusion 2-category $\e{C}$, its looping $\Omega\e{C}$ is defined as 
\begin{align}
    \Omega\e{C}:=\hom_{\e{C}}(\bbm{1},\bbm{1}),
\end{align}
where $\bbm{1}$ is the monoidal unit of $\e{C}$. This is always a braided fusion category. We may then examine its Müger center $\Z_2(\Omega\e{C})$, which is a symmetric fusion category. We say the fusion 2-category $\e{C}$ is bosonic if $\Z_2\Omega\e{C}$ is Tannakian(isomorphic to $\rep(G)$), and fermionic if $\Z_2\Omega\e{C}$ is super-Tannakian(isomorphic to $\rep(G,z)$ for some super group $(G,z)$). In this work, we are only concerned with bosonic fusion 2-categories.  In 2+1-D,  a general finite internal symmetry is described by a fusion 2-category. The \symto of a fusion 2-category is given by its $E_1$-center(also known as the Drinfeld center):
\begin{align}
    \Z_1[\e{C}]:=\edo_{\e{C}-\e{C}}(\e{C}),
\end{align}
which is a non-degenerate braided fusion 2-category, i.e. a 3+1D topological order. Some examples of fusion 2-categories that will appear in this work are listed below.

\begin{example}
    \begin{itemize}
    \item For any braided fusion category $\cc{B}$, we denote by $\Sigma\cc{B}:=\Mod(\cc{B})$ the 2-category of finite semi-simple left $\cc{B}$-module 2-categories. This is naturally a fusion 2-category with monoidal product given by relative tensor product over $\cc{B}$~\cite{douglas2018fusion2categoriesstatesuminvariant}.
    \item For any 2-group $\cc{G}$, the fusion 2-category of $\cc{G}$-graded finite semi-simple categories is denoted as $2\vc_{\cc{G}}$.  One can also twist the monoidal structure by any class $\omega\in \cc{H}^4[B\cc{G},\bC^\times]$, the resulting fusion 2-category is denoted as $2\vc_{\cc{G}}^\omega$.
    \item For any 2-group $\cc{G}$, the fusion 2-category of 2-representations of $\cc{G}$ is denoted as $2\rep(\cc{G})$. 
\end{itemize}
\end{example}

For any bosonic fusion 2-category $\e{C}$, its $E_1$-center must be a Dijkgraaf-Witten theory, i.e. a (potentially twisted) finite gauge theory, $\Z_1[2\vc_G^\pi]$~\cite{décoppet2024drinfeldcentersmoritaequivalence}. The complete braided fusion 2-category structure of $\Z_1[2\vc_G^\pi]$ was first studied in ~\cite{Kong_2020_center}. 

Fusion 2-categories have been  classified~\cite{décoppet2024classificationfusion2categories}. We summarize here the classification in the bosonic case.
\begin{fact}
    Bosonic fusion 2-categories are parameterized by
\begin{itemize}
    \item A finite group $G$.
    \item A class $\pi\in \cc{H}^4[G,U(1)]$.
    \item A subgroup $H<G$ up to conjugation.
    \item A non-degenerate braided fusion category $\cc{B}$ with $H$-action $\rho$.
    \item A class $\psi\in C^3[H,U(1)]$ such that $d\psi=\cc{O}_4(\rho)\pi|_H^{-1}$, where $\cc{O}_4(\rho)$ is the anomaly of the $H$-action $\rho$.
\end{itemize}
\end{fact}

The corresponding fusion 2-category will be denoted as $\fk{C}[G,H,\pi,\cc{C},\psi]$. Among these data, $G,\pi$ determines the \symto: $\Z_1[\fk{C}[G,H,\pi,\cc{C},\psi]]\simeq \Z_1[2\vc_G^\pi]$. In other words, any fusion 2-category of the form $\fk{C}[G,H,\pi,\cc{C},\psi]$ can be realized as a gapped boundary of $\Z_1[2\vc_G^\pi]$. Below we discuss how the examples discussed before fit into this classification.
\begin{example}\label{eg: fusion 2-cat_1}
 Let $\cc{B}$ be a braided fusion category. $\Sigma\cc{B}$ is a bosonic fusion 2-category if and only if $\Z_2\cc{B}$ is Tannakian. Say $\Z_2\cc{B}=\rep(G)$, then the de-equivariantization $\cc{B}_G$ is a non-degenerate braided fusion category with a natural $G$-action. Then we have $\Sigma\cc{B}\simeq\fk{C}[G,G,\pi,\cc{B}_G,\psi]$, where $\pi$ is cohomologous to the $\cc{O}_4-$anomaly of the $G$-action on $\cc{B}_G$ and $\psi$ is an arbitrary cochain such that $d\psi=\cc{O}_4\pi^{-1}$.
\end{example}
\begin{example}\label{eg: fusion 2-cat_2}
    Let $\cc{G}=A^{[0]}\rtimes H^{[1]}$ be a split finite 2-group, then $2\vc_{\cc{G}}\simeq \fk{C}[\widehat{A}\rtimes H,\widehat{A},1,\vc,1]$ and $2\rep(\cc{G})\simeq \fk{C}[\widehat{A}\rtimes H,H,0,\vc,0]$, where $\widehat{A}$ is the Pontryagin dual of $A$. The general(non-split) case can be found in~\cite{D_coppet_2025}.
\end{example}

\section{Topological holography for 2+1-D gapped phases\label{sec: symtft gapped}}
In this section we discuss generalities of topological holography for 2+1-D  gapped phases with fusion 2-categorical symmetries. In the gapped scenario, most constructions can be rigorously defined and computed.  We begin by reviewing the definition of 2+1D gapped phases with symmetry. We then formulate the sandwich construction in terms of relative tensor product. When the symmetry boundary and physical boundary are Morita equivalent, the sandwich admits a particularly simple description as a 3-representation of the symmetry 2-category.
\subsection{Gapped 2+1-D phases with symmetry}
A bosonic gapped 2+1-D phase is determined by its topological order up to stacking with the invertible $E_8$-state~\footnote{In the rest of this work, we will neglect the subtly caused by the invertible $E_8$-state and view gapped phase and topological order as synonymous.}. We recall the classification of  2+1-D topological orders:
\begin{fact}[\cite{Johnson_Freyd_2022}]
     A 2+1-D topological order(with possible local degeneracy) is a multi-fusion 2-category $\e{T}$ such that $\Z_1[\e{T}]=2\vc$.
\end{fact}
Physically, $\e{T}$ describes all the topological defects of the topological order and their fusion.  The condition $\Z_1[\e{T}]=2\vc$ is an anomaly-free condition: the topological order can be realized by a lattice model in pure 2+1-D. If there is no local degeneracy, then $\e{T}$ is fusion, and the anomaly-free condition implies that $\e{T}=\Sigma\cc{B}$ for some non-degenerate braided fusion category $\cc{B}$~\cite{Johnson_Freyd_2022}.

 We have now defined symmetries and gapped phases in 2+1-D. The definition of a 2+1-D gapped phase with symmetry is as simple as it gets.

\begin{definition}
    A 2+1-D gapped phase $\e{T}$ with symmetry $\e{C}$ is a (linear monoidal 2-)functor :
\begin{align}
    \e{F}: \e{C}\to \e{T}.
\end{align}
\end{definition}
To see the correctness of this definition, we recall some familiar examples.
\begin{example}
    \begin{enumerate}
    \item If the underlying topological order is trivial: $\e{C}=2\vc$, and the symmetry is a a finite group 0-form symmetry $2\vc_G$, then a functor $\e{F}: 2\vc_G\to 2\vc$ is exactly defined(up to equivalence) by a class $\omega\in\cc{H}^3[G,\bC^\times]$. We recover the cohomology classification of 2+1-D $G$-SPTs.
    \item More generally, if the symmetry is a non-anomalous 2-group symmetry $\e{C}=2\vc_{\cc{G}}$, then a linear monoidal functor $\e{F}: 2\vc_{\cc{G}}\to 2\vc$ is the same as a homotopy class of map $B\cc{G}\to B^3\bC^\times$, i.e. a class in $\cc{H}^3[\cc{G},\bC^\times]$. We recover the cohomology classification of 2+1-D $\cc{G}$-SPTs.
    \item If the underlying topological order has no local GSD, i.e. takes the form $\Sigma\cc{B}$, and the symmetry is a finite group 0-form symmetry $2\vc_G$, then a functor $\e{F}: 2\vc_G\to \Sigma\cc{B}$ is the same as a homotopy class of maps
    \begin{align}
        \e{F}: BG\to B\underline{\underline{\mathbf{Pic}}}(\cc{B})
    \end{align}
    where $\underline{\underline{\mathbf{Pic}}}(\cc{B})$ is the Picard 3-group of $\cc{B}$~\footnote{See Appendix~\ref{app: extension theory} for definition.}. Such maps are in 1-1 correspondence with $G$-crossed extensions of $\cc{B}$. This is exactly the classification of $G$-symmetry enriched topological orders(SET) in 2+1-D.
    \item More generally, if the finite group 0-form symmetry is anomalous, then a functor $\e{F}: 2\vc_G^\pi\to \Sigma\cc{B}$  describes a 2+1-D SET with an anomalous $G$-symmetry. 
\end{enumerate}
\end{example}

The examples considered above all have invertible symmetries. In general it is difficult to construct 2+1-D topological orders with non-invertible symmetries systematically. \symto provides an efficient method for achieving this. 
\subsection{The sandwich construction}
To construct phases with symmetry $\e{C}$ via  \symto, we start with the 3+1D topological order $\Z_1[\e{C}]$, and put it on a sandwich geometry with two boundaries, dubbed symmetry boundary, $\fk{B}_\phys$, and physical boundary $\fk{B}_\phys$. The topological order $\Z_1[\e{C}]$ has a canonical gapped boundary condition, where the topological operators form exactly $\e{C}$. This is chosen as the boundary condition at $\fk{B}_\sym$. Upon interval compactification, the sandwich reduces to a 2+1-D system with symmetry $\e{C}$. We may informally write this 2+1-D system as
\begin{align}
    \fk{B}_\phys\boxtimes_{\Z_1[\e{C}]}\fk{B}_\sym=\fk{B}_\phys\boxtimes_{\Z_1[\e{C}]}\e{C},\label{eq: sandwich}
\end{align}
and the symmetry $\e{C}$ simply acts by mapping into the second factor:
\begin{align}
    \e{F}: \e{C}\xrightarrow{\bbm{1}\boxtimes\id_\e{C}}\fk{B}_\phys\boxtimes_{\Z_1[\e{C}]}\e{C}.
\end{align}

However we have not given Eq.~\eqref{eq: sandwich} a precise mathematical meaning. For constructing gapped 2+1-D phases with symmetry $\e{C}$, we should consider $\fk{B}_\phys$ to be gapped as well. In this case, we may adopt the following description of gapped boundaries of 3+1D topological orders~\cite{kong2015boundarybulkrelationtopologicalorders,Kong_2017}: a gapped boundary of $\Z_1[\e{C}]$ is defined by a fusion 2-category $\e{D}$ together with a braided equivalence $\phi: \Z_1[\e{C}]\to \Z_1[\e{D}]$. This makes the gapped boundary $\e{D}$ into a so-called $E_1$-fusion module 2-category over $\Z_1[\e{C}]$. Then the expression Eq.~\eqref{eq: sandwich} is precisely the relative tensor product between left and right $E_1$-fusion module 2-categories over $\Z_1[\e{C}]$~\footnote{See Appendix~\ref{app: higher linear alg} for definition.}~\cite{kong2025highercondensationtheory}. Physically, the topological defects in the bulk can act on both the left and the right boundary. After interval compactification, these actions should be identified.

The dimension reduction of a gapped sandwich is particularly simple if $\fk{B}_\sym$ and $\fk{B}_\phys$ happen to be Morita equivalent~\footnote{Morita theory of fusion 2-categories is reviewed in Appendix~\ref{app: higher linear alg}}. In this case, we may write
\begin{align}
    \fk{B}_\phys=\e{C}_M^*:=\edo_{\e{C}}(\e{M})
\end{align}
for some (right)$\e{C}$-module 2-category $\e{M}$, and $\e{C}_M^*$ is the dual fusion 2-category with respect to $\e{M}$. Physically $\e{M}$ describes a gapped domain wall between $\fk{B}_\phys$ and $\fk{B}_\sym=\e{C}$. Then we have\hide{~\footnote{The general formula is 
\begin{align}
    \edo_{\e{C}-\e{D}}(\e{M})\boxtimes_{\edo_{\e{D}-\e{D}}(\e{D})}\edo_{\e{D}-\e{E}}(\e{N})=\edo_{\e{C}-\e{E}}(\e{M}\boxtimes_{\e{D}}\e{N}),
\end{align} 
where $\e{C},\e{D},\e{E}$ are fusion 2-categories,  $\e{M}$ is a $\e{C}-\e{D}$ bimodule 2-category, $\e{N}$ is a $\e{D}-\e{E}$ bimodule 2-category. See appendix~\ref{app: higher linear alg}.}}
\begin{align}
    \fk{B}_\phys\boxtimes_{\Z_1[\e{C}]}\fk{B}_\sym&=\edo_{\e{C}}(\e{M})\boxtimes_{\edo_{\e{C}-\e{C}}(\e{C})}\edo_{\e{C}}(\e{C})\nonumber\\
    &=\edo(\e{M}\boxtimes_{\e{C}}\e{C})=\edo(\e{M}).\label{eq: sandwich_2}
\end{align}

The symmetry $\e{C}$ then acts on $\edo(\e{M})$ according to the $\e{C}$-module structure on $\e{M}$:
\begin{align}
 \e{F}: \e{C}\to \edo(\e{M}), ~  a\mapsto -\lhd a.
\end{align}

Therefore the symmetry $\e{C}$ simply acts on the 3-vector space $\e{M}$ by endomorphism. This is a direct 3-categorification of 0+1D phases with symmetry, which are vector spaces with symmetry acting by endomorphism. This is  the most general form of a non-chiral 2+1-D topological order with symmetry.

\begin{figure}
    \centering
        \begin{tikzpicture}[baseline=(current  bounding  box.center)]
\draw[thick]  (0,0) -- (0,4);
\draw[thick]  (2,0) -- (2,4);
\fill[zima] (0,0) rectangle (2,4);
\node[label, below] at (0,0){$\fk{B}_\phys$};
\node[label,left] at (0,2){$\e{C}_{\e{M}}^*$};
\node[label, below] at (2.2,0){$\fk{B}_\sym$};
\node[label, right] at (2,2){$\e{C}$};
\node[label] at (1,2){$\Z_1[\e{C}]$};
\node[label] at (3,2){$\simeq$};
\filldraw[fill=zima, draw=black] (5,0) .. controls (6,2) .. (5,4) .. controls (4,2) .. (5,0);
\node[label, below] at (5,0){$\e{M}$};
\node[label, left] at (4,2){$\e{C}_{\e{M}}^*$};
\node[label,right] at (6,2){$\e{C}$};
\node[label] at (5,2){$\Z_1[\e{C}]$};
\node[label] at (7,2){$\simeq$};
\draw[thick] (8,0) -- (8,4);
\node[label, below] at (8,0){$\e{M}$};
\node[label, right] at (8,2){$\Z_0(\e{M})=\edo(\e{M})$};
\filldraw[fill=black] (5-0.1,0) rectangle (5+0.1,0.2);
\filldraw[fill=black] (5-0.1,4-0.2) rectangle (5+0.1,4);
\filldraw[fill=black] (8-0.1,0) rectangle (8+0.1,0.2);
\filldraw[fill=black] (8-0.1,4-0.2) rectangle (8+0.1,4);
\end{tikzpicture}
    \caption{If the physical and symmetry boundaries happen to be Morita equivalent, then we can bend them to make a a gapped domain wall $\e{M}$.  Therefore the dimension reduction of the sandwich is the bulk of the domain wall $\e{M}$. By ``bulk=center of boundary", we see that the dimension reduction of the sandwich must be $\Z_0(\e{M})=\edo(\e{M})$. The symmetry $\e{C}$ then acts on $\edo(\e{M})$ according to the module 2-category structure on $\e{M}$.}
    \label{fig: sandwich_2}
\end{figure}
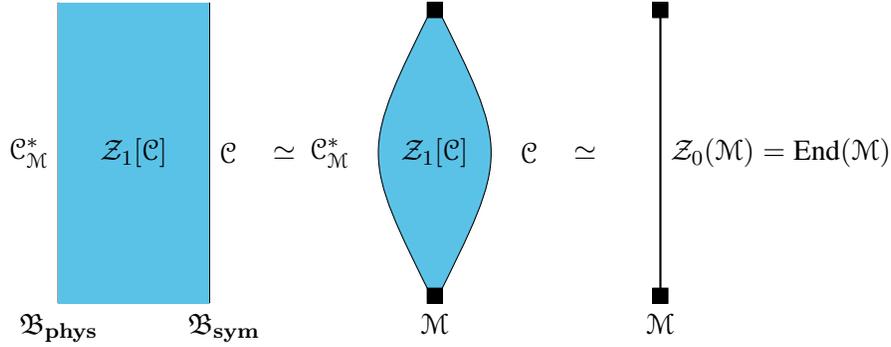

The interval compactification Eq.~\eqref{eq: sandwich_2} in fact directly follows from the bulk-boundary relation of 2+1-D topological orders: if the physical boundary is Morita equivalent to the symmetry boundary, then we can form a gapped domain wall $\e{M}$ between them, see Fig.~\ref{fig: sandwich_2}, then upon interval compactification the sandwich becomes the 2+1-D bulk of the 1+1-D boundary $\e{M}$. By the bulk-boundary relation in 2+1-D~\cite{kong2015boundarybulkrelationtopologicalorders,kong2025highercondensationtheory}, the sandwich must be the center of the boundary: $\Z_0(\e{M})=\edo(\e{M})$. We will study several examples of gapped phases obtained through this construction in the next section.

\section{2+1-D Gapped phases with generalized symmetries from \symto\label{sec: symtft gapped example}}
In this section we apply the \symto sandwich construction to study 2+1-D gapped phases with generalized symmetries. By our discussion in Sec.~\ref{sec: fusion 2-cat review}, possible gapped boundaries of $\Z_1[2\vc_G^\pi]$ are  fusion 2-categories of the form $\fk{C}[G,H,\pi,\cc{C},\psi]$. Here we focus on the case $\cc{C}=\vc$, and we will call the corresponding gapped boundary a minimal gapped boundary. Physically, such gapped boundaries have no intrinsic topological order. As we shall see, minimal gapped boundaries already give a large class of interesting 2+1-D gapped phases. We derive the general structure of a sandwich construction with minimal gapped symmetry and physical boundaries, including the underlying topological order and the symmetry action. We then study several concrete examples.
\subsection{Sandwiches with  minimal gapped boundaries\label{subsec: minimal}}
For a minimal gapped boundary of $\Z_1[2\vc_G^\pi]$, we will neglect the trivial factor $\cc{C}=\vc$ and denote the fusion 2-category of topological defects on the boundary as $\fk{C}[G,H,\pi,\psi]$. Such fusion 2-categories are called \textit{group-theoretical fusion 2-categories} in ~\cite{D_coppet_2025}. They cover a wide range of fusion 2-categories, including $2\vc_{\cc{G}}, 2\rep(\cc{G})$ for any 2-group $\cc{G}$. By the results in~\cite{D_coppet_2025}, $\fk{C}[G,H,\pi,\psi]$ is Morita equivalent to $2\vc_G^{\pi}$. Specifically, we have 
\begin{align}
    \fk{C}[G,H,\pi,\psi]\simeq \Bimod_{\vc_H^\psi-\vc_H^\psi}(2\vc_G^\pi).
\end{align}
where $\vc_H^\psi$ is viewed as a condensable $E_1$-algebra in $2\vc_G^\pi$~\footnote{For definition of condensable $E_1$-algebra, see Appendix~\ref{app: higher linear alg}.}. A more explicit description of this fusion 2-category is~\cite{D_coppet_2025}
\begin{align}
    &\fk{C}[G,H,\pi,\psi]=\mathop\boxplus_{[g]\in H\backslash G/H} \Sigma \vc_{H\cap \cg{g}{H}}^{\xi_g},\label{eq: symmetry_category}
\end{align}
where $H\backslash G/ H$ is the set of double cosets and $\xi_g$ is a 3-cocyle constructed from $\pi$ and $\psi$. In the following we will need the more general case,~\cite{D_coppet_2025}
\begin{align}
     &\Bimod_{\vc_H^\psi-\vc_K^\phi}(2\vc_G^\pi)=\mathop\boxplus_{[g]\in H\backslash G/ K}\Sigma \vc_{H\cap\cg{g}{K}}^{\xi_g},\label{eq: minimal_bimod}\\
     \xi_g(h_1, h_2, h_3) &:= \frac{\psi(h_1, h_2, h_3)}{\omega(g^{-1} h_3^{-1} g, g^{-1} h_2^{-1} g, g^{-1} h_1^{-1} g)} \notag \\
    &\quad \times \frac{\pi(h_3, h_3^{-1} g, g^{-1} h_2^{-1} g, g^{-1} h_1^{-1} g) \pi(g, g^{-1} h_3^{-1} g, g^{-1} h_2^{-1} g, g^{-1} h_1^{-1} g)}{\pi(h_2, h_3, h_3^{-1} h_2^{-1} g, g^{-1} h_1^{-1} g) \pi(h_1, h_2, h_3, h_3^{-1} h_2^{-1} g)}.\notag
\end{align}

Denote the 2-category of modules over $\vc_H^\psi$ in $2\vc_G^\pi$ as  $\e{M}_{H,\psi}:=\Mod_{\vc_{H}^\psi}(2\vc_G^\pi)$, we have
\begin{align}
    \edo_{2\vc_G^\pi}(\e{M}_{H,\psi})=\Bimod_{\vc_H^\psi-\vc_H^\psi}(2\vc_G^\pi).
\end{align}

Now consider a sandwich with a $\Z_1[2\vc_G^\pi]$ bulk, a symmetry boundary $\fk{C}[G,H,\pi,\psi]$ and a physical boundary $\fk{C}[G,K,\pi,\phi]$. Then the interval compactification of the sandwich can be computed as follows,
\begin{align}
    \e{T}&=\fk{B}_{\sym}\boxtimes_{\text{\symto}}\fk{B}_\phys\nonumber\\
    &=\fk{C}[G,H,\pi,\psi]\boxtimes_{\Z_1[2\vc_G^\pi]}\fk{C}[G,K,\pi,\phi]\nonumber\\
    &=\edo_{2\vc_G^\pi}(\e{M}_{H,\psi})\boxtimes_{\Z_1[2\vc_G^\pi]} \edo_{2\vc_G^\pi}(\e{M}_{K,\phi}).
\end{align}

Using Eq.~\eqref{eq: app_relative_E1} and then Eq.~\eqref{eq: app_relative}, this is equal to 
\begin{align}
    \e{T}&=\edo(\e{M}_{H,\psi}\boxtimes_{2\vc_G^\pi}\e{M}_{K,\phi})\label{eq: minimal_1}\\
    &=\edo(\Bimod_{\vc_H^\psi-\vc_K^\phi}(2\vc_G^\pi))\label{eq: minimal_2}
\end{align}

Now using Eq.~\eqref{eq: minimal_bimod}, this is equal to
\begin{align}
   \e{T} &=\edo(\mathop\boxplus_{[g]\in H\backslash G/K} \Sigma\vc^{\xi_g}_{H\cap \cg{g}{K}})\nonumber\\
    &=\mathop\boxplus_{[g_1],[g_2]}\mathbf{Func}[\Sigma\vc^{\xi_{g_1}}_{H\cap \cg{g_1}{K}}, \Sigma\vc^{\xi_{g_2}}_{H\cap \cg{g_2}{K}}]\nonumber\\
    &=\mathop\boxplus_{[g_1],[g_2]}\Bimod_{\vc^{\xi_{g_1}}_{H\cap \cg{g_1}{K}}-\vc^{\xi_{g_2}}_{H\cap \cg{g_2}{K}}}(2\vc)\label{eq: sandwich_minimal}
\end{align}

From this, we can read off the structure of the underlying topological order of the sandwich:
\begin{itemize}
    \item In general it has multiple vacuua. This happens when the double coset $H\backslash G/K$ contains more than one element, which means $HK\neq G$. Physically, the local vacuum degeneracy comes from spontaneous breaking of 0-form symmetries. 
    The number of vacuua is equal to the cardinality of the double coset
    \begin{align}
     \mathbf{GSD}=|H\backslash G/K|.
    \end{align}
    \item Each vacuum has topological order 
    \begin{align}
       \e{T}_{[g],[g]}:=\Bimod_{\vc^{\xi_{g}}_{H\cap \cg{g}{K}}-\vc^{\xi_{g}}_{H\cap \cg{g}{K}}}(2\vc)=\Sigma\Z_1[\vc^{\xi_g}_{H\cap \cg{g}{K}}],
    \end{align}
    i.e. a 2+1-D $H\cap \cg{g}{K}$-gauge theory with twist $\xi_g$. Since there is no intrinsic topological order on the symmetry or the physical boundary, the topological order of the sandwich comes from spontaneous breaking of (possibly non-invertible)1-form symmetries.
    \item The 2-category of domain walls between vacuum $[g]$ and vacuum $[h]$ is 
    \begin{align}
     \e{T}_{[g],[h]}:=\Bimod_{\vc^{\xi_{g}}_{H\cap \cg{g}{K}}-\vc^{\xi_{h}}_{H\cap \cg{h}{K}}}(2\vc)
    \end{align}
    
    This is exactly the 2-category of ALL domain walls between $\Z_1[\vc^{\xi_g}_{H\cap \cg{g}{K}}]$ and $\Z_1[\vc^{\xi_h}_{H\cap \cg{h}{K}}]$~\cite{Kitaev_2012}.
    \item The fusion rule of domain walls is given by taking relative tensor product: for $\cc{M}_{[g],[h]}\in \e{T}_{[g],[h]}$, $\cc{M}_{[h],[k]}\in\e{T}_{[h],[k]}$, we have 
    \begin{align}
        \cc{M}_{[g],[h]}\otimes \cc{M}_{[h],[k]}:=\cc{M}_{[g],[h]}\boxtimes_{\vc_{H\cap \cg{h}{K}}^{\xi_h}}\cc{M}_{[h],[k]}.
    \end{align}
\end{itemize}

The symmetry category is $\e{C}=\Bimod_{\vc_H^\psi-\vc_H^\psi}(2\vc_G^\pi)$, the symmetry action functor $\e{F}: \e{C}\to \e{T}$ is given by relative tensor product: For any symmetry generator $\cc{D}\in \e{C}$,
\begin{align}
   \e{F}(\cc{D})=\cc{D}\boxtimes_{\vc_{H}^\psi}- \in \edo(\Bimod_{\vc_H^\psi-\vc_K^\phi}(2\vc_G^\pi)).\label{eq: sym_act}
\end{align}
\subsection{Anomaly of fusion 2-categorical symmetry}
The above analysis allows us to  determine whether a general fusion 2-categorical symmetry is anomalous. Recall that a fusion 2-categorical symmetry $\e{C}$ is called non-anomalous if there is exists a trivial gapped phase with symmetry $\e{C}$. This means there is a fiber 2-functor 
\begin{align}
    \e{C}\to 2\vc.
\end{align}

Using the sandwich construction, this means there exists a gapped physical boundary of $\Z_1[\e{C}]$, such that the sandwich is a trivial 2+1-D topological order with no local degeneracy. If the symmetry boundary itself has intrinsic 2+1-D topological order, then the sandwich must have nontrivial topological order. Therefore for $\e{C}$ to be non-anomalous, the symmetry boundary itself must be a minimal gapped boundary. Similarly the physical boundary must be a minimal gapped boundary for the sandwich to have trivial topological order. Then by our analysis before, the sandwich is a trivial topological order if and only if 
\begin{enumerate}
    \item $\mathbf{GSD}=|H\backslash G/K|=1$, meaning that $G=HK$.
    \item The topological order $\Z_1[\vc_{H\cap K}^{\xi_e}]$ is trivial, meaning that $H\cap K=\{e\}$.
\end{enumerate}

This leads to the following.
\begin{corollary}
    A bosonic fusion 2-categorical symmetry $\fk{C}[G,H,\cc{B},\pi,\psi]$ is non-anomalous if and only $\cc{B}=\vc$, and there exists $K<G$ such that $H\cap K=\{e\}$ and $HK=G$.
\end{corollary}
This result is known in mathematics literature~\cite{D_coppet_2025}. 
\begin{example}\label{eg: anomalous 2-group}
    Let $G=\bZ_4,H=\bZ_2$, then the symmetry $\fk{C}[\bZ_4,\bZ_2,0,\vc,0]$ is anomalous. In fact it is equivalent to the anomalous 2-group symmetry 
    \begin{align}
        2\vc_{\bZ_2^{[0]}\times\bZ_2^{[1]}}^\omega
    \end{align}
    with a nontrivial class $\omega\in \cc{H}^4[\bZ_2^{[0]}\times\bZ_2^{[1]},\bC^\times]$~\cite{D_coppet_2025}.
\end{example}
\subsection{Example: Gapped phases from $\Z_1[2\vc_{\bZ_2}]$}
Topological holography with \symto $\Z_1[2\vc_{\bZ_2}]$ has been studied in~\cite{bhardwaj2024gappedphases21dnoninvertible}. Although this example is relatively simple, it serves to illustrate the general strategy.

For $\Z_1[2\vc_{\bZ_2}]$, there are three minimal gapped boundaries: 
\begin{itemize}
    \item One with $H=\{1\}$. This is often called the rough boundary, denoted as $\fk{B}_e$.
    \item One with $H=\bZ_2$ and $\psi=0$. This boundary condenses the flux $m$. This is often called the smooth boundary, denoted as $\fk{B}_m$.
    \item One with $H=\bZ_2$ and a nontrivial twist $\psi \in\cc{H}^3[\bZ_2,U(1)]$. This boundary also condenses the flux, but the endpoint of a flux is a semion~\cite{Zhao_2023}. This is called the twisted smooth boundary, denoted as $\fk{B}_m^\psi$.
\end{itemize}
\subsubsection{$\fk{B}_\sym=\fk{B}_e$, $\fk{B}_\phys=\fk{B}_e$}
This corresponds to $H=K=\{1\}$. The double coset $H\backslash \bZ_2/K=\{[1],[g]\}$. By Eq.~\eqref{eq: symmetry_category}, the symmetry category is 
\begin{align}
    \e{C}=\mathop\boxplus_{1,g}\Sigma\vc=2\vc_{\bZ_2},
\end{align}
a 0-form $\bZ_2$-symmetry. By Eq.~\eqref{eq: sandwich_minimal}, the underlying topological order of the sandwich is 
\begin{align}
    \e{T}=\edo(\mathop\boxplus_{1,g}\Sigma\vc)=\mathop\boxtimes_{i,j=1,g}\Sigma\vc.
\end{align}

In other words, there are two vacuua, each is in a trivial gapped phase. In the symmetry category $2\vc_{\bZ_2}$, there is only one nontrivial symmetry generator, denoted as  $\cc{D}_g$, with $\cc{D}_g\otimes \cc{D}_g=1$. By Eq.~\eqref{eq: sym_act}, it is mapped to the operator 
\begin{align}
  \e{F}(\cc{D}_g)=\vc_g\boxtimes-= \begin{pmatrix}
        0& \vc\\
        \vc &0
    \end{pmatrix}\in \edo(\mathop\boxplus_{1,g}\Sigma\vc).
\end{align}

 We have $\e{F}(\cc{D}_g)(\vc_1)=\vc_g, \e{F}(\cc{D}_g)(\vc_g)=\vc_1$. In other words, the two vacuua are exchanged under the action of the $\bZ_2$-symmetry. We conclude that this sandwich describes a $\bZ_2$-SSB gapped phase.
\subsubsection{$\fk{B}_\sym=\fk{B}_m$, $\fk{B}_\phys=\fk{B}_m$}
This corresponds to $H=K=\bZ_2$, the double coset $H\backslash G/K=\{[1]\}$ and $H\cap K=\bZ_2$. The symmetry boundary is described as follows.
\begin{itemize}
    \item By Eq.~\eqref{eq: symmetry_category}, the symmetry category is 
\begin{align}
    \e{C}=\Sigma\vc_{\bZ_2}\simeq\Sigma\rep(\bZ_2).
\end{align}
i.e. a $\bZ_2$ 1-form symmetry.
\item  It has two simple objects($\rep(\bZ_2)$-module categories),\footnote{Generally an indecomposable $\rep(G)$-module category is labeled by a subgroup $H<G$ and a 2-cocyle $\psi$ of $H$. We denote such a module category by $\cc{D}^{H,\psi}$. }
\begin{align}
    \cc{D}^{\bZ_2}:=\rep(\bZ_2), ~ \cc{D}^{0}:=\vc.
\end{align}
\item The fusion rule is 
\begin{align}
    &\rep(\bZ_2)\boxtimes_{\rep(\bZ_2)}\rep(\bZ_2)=\rep(\bZ_2),\nonumber\\
    &\rep(\bZ_2)\boxtimes_{\rep(\bZ_2)}\vc=\vc\nonumber\\
    &\vc\boxtimes_{\rep(\bZ_2)}\vc=\vc\mathop\boxplus \vc.
\end{align}
\end{itemize}

By Eq.~\eqref{eq: sandwich_minimal}, the underlying topological order of the sandwich is 
\begin{align}
    \e{T}=\edo(\Sigma\rep(\bZ_2))=\Sigma(\rep(\bZ_2)\boxtimes \rep(\bZ_2))=\Sigma\Z_1[\vc_{\bZ_2}],
\end{align}
i.e. a toric code.

\begin{figure}
    \centering
            \begin{tikzpicture}[baseline=(current  bounding  box.center)]
\draw[very thick]  (1,0) -- (1,3);
\draw[very thick] (1.2,0) -- (1.2,3);
\fill[zima] (0,0) rectangle (1,3);
\fill[zima] (1.2,0) rectangle (2.2,3);
\node[label, below] at (1.2,0){$\cc{M}_{m,m}$}; 
\end{tikzpicture}
\hspace{0.2cm}
               \begin{tikzpicture}[baseline=(current  bounding  box.center)]
\draw[very thick]  (1,0) -- (1,3);
\draw[very thick,dashed] (1.2,0) -- (1.2,3);
\fill[zima] (0,0) rectangle (1,3);
\fill[zima] (1.2,0) rectangle (2.2,3);
\node[label, below] at (1.2,0){$\cc{M}_{m,e}$}; 
\end{tikzpicture}
\hspace{0.2cm}
               \begin{tikzpicture}[baseline=(current  bounding  box.center)]
\draw[very thick,dashed]  (1,0) -- (1,3);
\draw[very thick] (1.2,0) -- (1.2,3);
\fill[zima] (0,0) rectangle (1,3);
\fill[zima] (1.2,0) rectangle (2.2,3);
\node[label, below] at (1.2,0){$\cc{M}_{e,m}$}; 
\end{tikzpicture}
\hspace{0.2cm}
               \begin{tikzpicture}[baseline=(current  bounding  box.center)]
\draw[very thick,dashed]  (1,0) -- (1,3);
\draw[very thick,dashed] (1.2,0) -- (1.2,3);
\fill[zima] (0,0) rectangle (1,3);
\fill[zima] (1.2,0) rectangle (2.2,3);
\node[label, below] at (1.2,0){$\cc{M}_{e,e}$}; 
\end{tikzpicture}
\hspace{0.2cm}
    \caption{Domain walls in the toric code topological order. A solid line represents a smooth boundary and a dashed line represents a rough boundary. There is also an invertible domain wall corresponding to a non-trivial braided auto-equivalence. As module category over $\rep(\bZ_2)$, the rough boundary is $\vc$ and the smooth boundary is $\rep(\bZ_2)$.}
    \label{fig: tc_dw}
\end{figure}
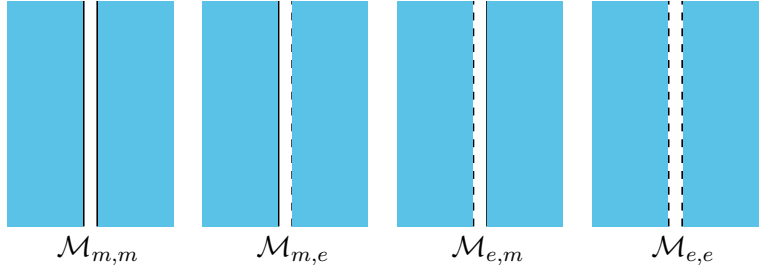

Before we describe the symmetry action on the toric code, let us recall the domain walls in a toric code, their description as $\rep(\bZ_2)-\rep(\bZ_2)$ bimodule categories, and their fusion rules.
\begin{itemize}
    \item A trivial domain wall $\cc{M}_1$, the bimodule category is $\rep(\bZ_2)$.
    \item Four domain walls corresponding to combinations of smooth/rough boundaries. See Fig.~\ref{fig: tc_dw}. They correspond to the bimodule categories
 $\cc{M}_{i,j}=N_i\boxtimes N_j$, with $i,j=e,m$ and $ N_e=\vc,N_m=\rep(\bZ_2)$. They have fusion rule
    \begin{align}
        \cc{M}_{i,j}\otimes \cc{M}_{k,l}=(\delta_{j,k}+1)\cc{M}_{i,l}
    \end{align}
    \item  An automorphism domain wall $\cc{M}_{\phi}$. This domain wall will not enter our discussion below.
\end{itemize}

By Eq.~\eqref{eq: sym_act}, the nontrivial symmetry generator $\cc{D}^{0}$ acts as
\begin{align}
    \e{F}(\cc{D}^{0})=\vc\boxtimes_{\rep(\bZ_2)}-\in \edo(\Sigma\rep(\bZ_2))\simeq \Sigma(\rep(\bZ_2)\boxtimes \rep(\bZ_2))
\end{align}

The isomorphism $\mathbf{Func}[\Sigma\e{A},\Sigma\e{B}]\simeq \Sigma(\e{A}^{op}\boxtimes \e{B})$ sends $f$ to $f(\e{A})$. Therefore as a $\rep(\bZ_2)-\rep(\bZ_2)$ bimodule category we have 
\begin{align}
    \e{F}(\cc{D}^0)=\vc\boxtimes_{\rep(\bZ_2)}\rep(\bZ_2)=\vc=\cc{M}_{e,e}.
\end{align}

We see that the symmetry generator $\cc{D}^{0}$ acts as the rough-rough domain wall of the toric code, also known as the Cheshire string. This domain wall is the condensation descendant of the gauge charge. The symmetry action is simply an embedding
\begin{align}
    \e{F}: \Sigma\rep(\bZ_2)\to \Sigma\Z_1(\vc_{\bZ_2})
\end{align}
induced by the embedding of $\rep(\bZ_2)\to \Z_1[\vc_{\bZ_2}]$. In other words, the 1-form $\bZ_2$-symmetry acts as the $e$-particle of the toric code. The 1-form symmetry is broken since there is an order parameter-- the $m$-particle.
\subsection{Example: Gapped phases from $\Z_1[2\vc_{S_3}]$}
We now consider a \symto $\Z_1[2\vc_{S_3}]$, where $S_3=\bZ_3\rtimes \bZ_2=\<r,s|r^3=s^2=1, srs=r^2\>$. This example is particularly interesting for the following reasons.
\begin{itemize}
    \item It admits a symmetry boundary describing a non-invertible fusion 2-category symmetry $2\rep(\bZ_3^{[1]}\rtimes \bZ_2^{[0]})$. 
    \item This non-invertible symmetry admits a symmetry-broken phase where different vacuua have distinct topological orders.
    \item This non-invertible symmetry itself is non-anomalous and admits SPT phases.
\end{itemize}

There are two non-trivial subgroups(up to conjugation): $\bZ_3=\<r\>$ and $\bZ_2=\<s\>$. Let $\fk{C}[S_3,H,\pi=0,\psi=0]$ be the symmetry boundary, and $\fk{C}[S_3,K,\pi=0,\phi=0]$ be the physical boundary. With $H, K$ being either the $\bZ_3$ or the $\bZ_2$ subgroup of $S_3$, we obtain four sandwiches. In the following we study these four cases. 

\subsubsection{$H=K=\bZ_3$}
We have $H\backslash G/K=\{[1],[s]\}$ and $H\cap K=\bZ_3$. The symmetry category is described as follows.
\begin{itemize}
    \item By Eq.~\eqref{eq: symmetry_category}, the underlying 2-category is 
    \begin{align}
        \e{C}=\Sigma\vc_{\bZ_3,[1]}\mathop\boxplus \Sigma\vc_{\bZ_3,[s]}\simeq\Sigma\rep(\bZ_3)_{[1]}\mathop\boxplus \Sigma\rep(\bZ_3)_{[s]}.
    \end{align}
    \item There are two connected components, each isomorphic to $\Sigma\rep(\bZ_3)$. This means the symmetry is necessarily a 2-group symmetry~\cite{D_coppet_2025}. Notice that conjugating with $s$ exchanges the two non-trivial irreducible representations of $\bZ_3$, therefore this is a 2-group symmetry
    \begin{align}
        \e{C}=2\vc_{\bZ_3^{[1]}\rtimes \bZ_2^{[0]}}.
    \end{align}
    where the 0-form $\bZ_2$-symmetry acts nontrivially on the 1-form $\bZ_3$-symmetry\footnote{This also follows from the general case Example.~\ref{eg: fusion 2-cat_2}}.
    \item There are four simple objects, two in each connected component, denoted as $\cc{D}_{[1]}^{\bZ_3}\simeq  \rep(\bZ_3),\cc{D}^{0}_{[1]}\simeq \vc, \cc{D}_{[s]}^{\bZ_3}\simeq \rep(\bZ_3), \cc{D}_{[s]}^0\simeq \vc$. The unit is $\cc{D}_{[1]}^{\bZ_3}$. The fusion rules are
    \begin{align}
    \begin{split}
        &\cc{D}_{[1]}^0\otimes \cc{D}_{[1]}^0=3\cc{D}_{[1]}^0,\\
        &\cc{D}_{[1]}^0\otimes \cc{D}_{[s]}^{\bZ_3}=\cc{D}^{0}_{[s]},\\
        &\cc{D}^{0}_{[1]}\otimes \cc{D}^{0}_{[s]}=3\cc{D}_{[s]}^0,\\
        &\cc{D}_{[s]}^{\bZ_3}\otimes \cc{D}_{[s]}^{\bZ_3}=\cc{D}^{\bZ_3}_{[1]},\\
        &\cc{D}_{[s]}^{\bZ_3}\otimes \cc{D}_{[s]}^0=\cc{D}_{[1]}^0,\\
        &\cc{D}_{[s]}^0\otimes \cc{D}_{[s]}^0=3\cc{D}_{[1]}^0.
    \end{split}\label{eq: 2-group_fusion}
    \end{align}
\end{itemize}

By Eq.~\eqref{eq: sandwich_minimal}, the underlying topological order of the sandwich is 
\begin{align}
    \e{T}&=\edo(\e{C})\\
    &=\mathop\boxplus_{i,j=[1],[s]}\Sigma(\rep(\bZ_3)\boxtimes\rep(\bZ_3)).
\end{align}

In other words, we have two vacuua, each supporting a $\bZ_3$-toric code. Next we analyze the symmetry action on this topological order. 

Consider the symmetry generator $\cc{D}_{[1]}^0$, we have 
\begin{align}
    \e{F}(\cc{D}_{[1]}^0)=\cc{D}_{[1]}^0\otimes-\in \edo(\Sigma\rep(\bZ_3)_{[1]}\mathop\boxplus \Sigma\rep(\bZ_3)_{[s]})
\end{align}
Since the isomorphism $\mathbf{Func}[\Sigma\e{A},\Sigma\e{B}]\simeq \Sigma(\e{A}^{op}\boxtimes \e{B})$ sends $f$ to $f(\e{A})$, to find the matrix elements $\e{F}(\cc{D}_{[1]}^0)_{i,j},~i,j=[1],[s]$, we only need to compute the action of $\e{F}(\cc{D}_{[1]}^0)$ on the ``basis" $\rep(\bZ_3)_{[1]}\simeq \cc{D}_{[1]}^{\bZ_3}, \rep(\bZ_3)_{[s]}\simeq \cc{D}_{[s]}^{\bZ_3}$. Using the fusion rules Eq.~\eqref{eq: 2-group_fusion}, we have 
\begin{align}
   & \e{F}(\cc{D}_{[1]}^0)(\cc{D}_{[1]}^{\bZ_3})=\cc{D}_{[1]}^0=\vc,\\
    & \e{F}(\cc{D}_{[1]}^0)(\cc{D}_{[s]}^{\bZ_3})=\cc{D}_{[s]}^0=\vc.
\end{align}
Therefore we have
\begin{align}
    \e{F}(\cc{D}_{[1]}^0)=\begin{pmatrix}
        \vc & 0\\
        0&\vc
    \end{pmatrix}\in \edo(\Sigma\rep(\bZ_3)\mathop\boxplus \Sigma\rep(\bZ_3)).
\end{align}

Similar to the $\bZ_2$-toric code, the  $\rep(\bZ_3)-\rep(\bZ_3)$-bimodule category $\vc$ corresponds to the rough-rough domain wall in the $\bZ_3$-gauge theory. Therefore 
\hide{$\pi_0\e{C}=\Sigma\rep(\bZ_3)$ simply embeds into the diagonal of the topological order:
\begin{align}
    \e{F}: \pi_0\e{C}=\Sigma\rep(\bZ_3)\to \Sigma\Z_1[\rep(\bZ_3)]\xrightarrow{\text{diag}}\edo(\Sigma\rep(\bZ_3)\mathop\boxplus \Sigma\rep(\bZ_3)).
\end{align}

In other words,} the 1-form $\bZ_3$-symmetry acts as the $e$-particle of the $\bZ_3$-gauge theory in each vacuum.
Next we consider the symmetry generator $\cc{D}_{[s]}^{\bZ_3}$, using fusion rules of $\e{C}$, it is not hard to deduce
\begin{align}
    \e{F}(\cc{D}_{[s]}^{\bZ_3})=\begin{pmatrix}
        0 & \rep(\bZ_3)\\
        \rep(\bZ_3) &0
    \end{pmatrix}\in \edo(\Sigma\rep(\bZ_3)\mathop\boxplus \Sigma\rep(\bZ_3)).
\end{align}

The $\rep(\bZ_3)-\rep(\bZ_3)$-bimodule category $\rep(\bZ_3)$ is the trivial domain wall between two $\Z_1[\rep(\bZ_3)]$. Therefore $\cc{D}_{[s]}^{\bZ_3}$ simply exchanges the two vacuua-- the 0-form $\bZ_2$-symmetry is broken.
\subsubsection{$H=K=\bZ_2$}
We have $H\backslash G/ K=\{[1],[r]\}$ and $H\cap K=\bZ_2$. Notice $\bZ_2$ is not normal in $S_3$ and $H\cap\cg{[r]}{K}=\{1\}$. The symmetry category is described as follows.
\begin{itemize}
    \item By Eq~\eqref{eq: symmetry_category}, the underlying 2-category is 
    \begin{align}
        \e{C}=\Sigma\vc_{\bZ_2}\mathop\boxplus \Sigma\vc=\Sigma\rep(\bZ_2)_{[1]}\mathop\boxplus \Sigma\vc_{[r]}.
    \end{align}
    \item There are two connected components, however they are not isomorphic. This means the symmetry is not a 2-group symmetry. 
    \item There are three simple objects, two in $\Sigma\rep(\bZ_2)_{[1]}$ and one in $\Sigma\vc_{[r]}$, denoted as $\cc{D}^{\bZ_2}\simeq \rep(\bZ_2),\cc{D}^{0}\simeq \vc,\cc{W}\simeq \vc$. $\cc{D}^{\bZ_2}$ is the unit, and fusion rules of other objects are~\cite{Bhardwaj_2023_2,D_coppet_2025}
    \begin{align}
    \begin{split}
        &\cc{D}^{0}\otimes \cc{D}^{0}=2\cc{D}^{0},\\
        &\cc{D}^{0}\otimes \cc{W}=2\cc{W},\\
        &\cc{W}\otimes \cc{W}=\cc{W}\oplus \cc{D}^{0}.
    \end{split}
        \label{eq: 2rep_1}
    \end{align}
    which we recognize is the fusion 2-category $2\rep(\bZ_3^{[1]}\rtimes \bZ_2^{[0]})$~\footnote{This follows from the general case Example~\ref{eg: fusion 2-cat_2}.}.
\end{itemize}

The underlying topological order of the sandwich is 
\begin{align}
    \e{T}=\edo(\Sigma\rep(\bZ_2)\mathop\boxplus \Sigma\vc).
\end{align}

 There are two vacuua, one has the toric code topological order, the other has trivial topological order. Next we analyze the action of $2\rep(\bZ_3^{[1]}\rtimes \bZ_2^{[0]})$ on this topological order. Consider the symmetry generator, $\cc{D}^{\bZ_2}$, it acts 
\begin{align}
    \e{F}(\cc{D}^{0})=\cc{D}^{0}\otimes -.
\end{align}

Using the fusion rules Eq.~\eqref{eq: 2rep_1}, we obtain

\begin{align}
    \e{F}(\cc{D}^{0})=\begin{pmatrix}
        _{\rep(\bZ_2)}\vc_{\rep(\bZ_2)} & 0\\
        0 &  _\vc (\vc\boxplus \vc)_\vc,
    \end{pmatrix}\in \edo(\Sigma\rep(\bZ_2)\mathop\boxplus \Sigma\vc),
\end{align}
where we added subscripts to express the module category structure. We see that in the toric code sector $\cc{D}^{\bZ_2}$ acts as the Cheshire string, while in the trivial phase sector it creates a 1+1D phase with two GSD, $2\bbm{1}$. Next we consider the symmetry generator $\cc{W}$, using the fusion rules Eq.~\eqref{eq: 2rep_1}, we obtain
\begin{align}
    \e{F}(\cc{W})=\begin{pmatrix}
        0 & _{\rep(\bZ_2)}\vc_\vc\\
        _\vc \vc_{\rep(\bZ_2)}& _\vc\vc_\vc
    \end{pmatrix}\in \edo(\Sigma\rep(\bZ_2)\mathop\boxplus \Sigma\vc).
\end{align}

One can directly verify, using relative tensor product of module categories, that the fusion rules Eq.~\eqref{eq: 2rep_1} are preserved under $\e{F}$.
\subsubsection{$H=\bZ_3, K=\bZ_2$}
Since $HK=S_3$ and $H\cap K=1$, we see from Eq.~\eqref{eq: sandwich}, that the underlying topological order of the sandwich is trivial($2\vc$). This means the sandwich is now a 2+1-D SPT with 2-group symmetry $2\vc_{\bZ_3^{[1]}\rtimes \bZ_2^{[0]}}$. There are in fact two SPTs, as the physical boundary can be twisted by a class $\phi\in\cc{H}^3[\bZ_2,U(1)]$. 

\begin{remark}
    Since these two SPTs differ by the twist $\phi\in\cc{H}^3[K,\bC^\times]$ on the physical boundary, they can be viewed as gauged version of ordinary(group-like) SPTs. In other words, if we change the symmetry boundary to $2\vc_G$, they would become different SPTs of the group-like symmetry $2\vc_K$.  In this sense they are not ``intrinsically non-invertible SPTs".
\end{remark}

\subsubsection{$H=\bZ_2, K=\bZ_3$}
Since $HK=S_3$ and $H\cap K=1$, we see from Eq.~\eqref{eq: sandwich} that the underlying topological order of the sandwich is trivial. Therefore the sandwich describes a 2+1-D SPT with non-invertible symmetry $2\rep(\bZ_3^{[1]}\rtimes \bZ_2^{[0]})$. There are in fact three SPTs, since the physical boundary can be twisted by a class $\phi\in\cc{H}^3[\bZ_3,U(1)]$.

 Similar to the previous case, these SPTs are not intrinsically non-invertible SPTs.

\subsection{Example: Gapped phases from $\Z_1[2\vc_{S_4}]$}
We now consider a \symto $\Z_1[2\vc_{S_4}]$, with $S_4=\<(1,2),(1,2,3,4)\>$ be the permutation group of four elements. This case is particularly interesting for the following reasons.
\begin{itemize}
    \item There exists a symmetry boundary that defines a non-invertible symmetry $\fk{C}[S_4,S_3,0,0]$ that is neither a 2-group symmetry $2\vc_{\cc{G}}$ nor a 2-representation symmetry $2\rep(\cc{G})$. In this sense this symmetry is  `` intrinsically non-invertible".
    \item The symmetry $\fk{C}[S_4,S_3,0,0]$ admits a partial symmetry-broken phase with two vacuua, with one vacuum having the $S_3$-gauge theory and the other having the $\bZ_2$-gauge theory. 
    \item The symmetry $\fk{C}[S_4,S_3,0,0]$ admits two families of SPT phases that are not related by twisting the physical boundary by cocycles as in the previous examples. Therefore it is impossible to gauge-relate these SPTs to a family of invertible SPTs with fixed symmetry group. These two families of SPTs are therefore ``intrinsically non-invertible SPTs". For instance, the gapless edge modes between these two families of SPTs should be distinct from the edge modes of any invertible SPTs.
\end{itemize}

\subsubsection{$H=K=S_3$}
We take $H=K=S_3=\<(1,2),(1,2,3)\>$. There are two double cosets, $H\backslash G/K=\{[e],[(3,4)]\}$. We have $H\cap \cg{(3,4)}{K}=\{e,(1,2)\}\simeq \bZ_2.$ The symmetry category is described as follows.
\begin{itemize}
    \item By Eq.~\eqref{eq: symmetry_category}, the underlying 2-category is 
    \begin{align}
        \e{C}=\Sigma\rep(S_3)_{e}\boxplus\Sigma\rep(\bZ_2)_{[(3,4)]}
    \end{align}
    \item Since $S_4$ is not a semi-direct product, by Example~\ref{eg: fusion 2-cat_2}, this is not a 2-group symmetry nor a 2-representation symmetry.
    \item There are in total six simple objects, four in $\Sigma\rep(S_3)_{e}$, two in $\Sigma\rep(\bZ_2)_{[(3,4)]}$. Let $g:=(3,4)$, the simple objects are denoted as 
    \begin{align}
        \D^{S_3}_e,\D^{\bZ_2}_e,\D^{\bZ_3}_e,\D^{0}_e,\D^{\bZ_2}_g,\D^{0}_{g}.
    \end{align}
    \item The fusion rules are~\footnote{This can be derived by decomposing $(H,K)$-bisets into unions of double cosets.}
    \begin{table}[h]
    \centering\small
    \def\arraystretch{2}
    \begin{tabular}{|C{1cm}|C{1cm}|C{1cm}|C{1cm}|C{1cm}|C{2cm}|C{2cm}|}
    \hline
      &$\D^{S_3}_e$ & $\D^{\bZ_2}_e$ &$\D^{\bZ_3}_e$ & $\D^0_e$ &$\D^{\bZ_2}_g$ &$\D^0_g$ \\
       \hline
       $\D^{S_3}_e$&$\D^{S_3}_e$ & $\D^{\bZ_2}_e$ &$\D^{\bZ_3}_e$ & $\D^0_e$ &$\D^{\bZ_2}_g$ &$\D^0_g$\\
        \hline
       $\D^{\bZ_2}_e$&&$3\D^{\bZ_2}_e$&$\D_e^0$&$3\D^0_e$&$3\D_g^{\bZ_2}$&$3\D^0_g$\\
       \hline
       $\D^{\bZ_3}_e$ &&&$2\D^{\bZ_3}_e$&$2\D^0_e$&$\D^0_g$&$2\D^0_g$\\
        \hline
       $\D^0_e$&&&&$6\D^0_e$&$3\D^0_g$&$6\D^0_g$\\
        \hline
        $\D^{\bZ_2}_g$ &&&&&$\D_e^{\bZ_2}\boxplus \D_g^0$&$\D_e^0\boxplus 2\D_g^0$\\
        \hline
       $\D^0_g$& &&&&&$2\D^0_e\boxplus4\D^0_g$\\
        \hline
    \end{tabular}
    \caption{Fusion rules of objects in $\fk{C}[S_4,S_3,0,0]$}
    \label{table: S4-S3 fusion}
\end{table}
\end{itemize}

The underlying topological order is, by Eq~\eqref{eq: sandwich_minimal},
\begin{align}
    \e{T}=\edo(\Sigma\rep(S_3)\boxplus \Sigma\rep(\bZ_2))
\end{align}

There are two vacuua, one has the $S_3$-gauge theory and the other has the toric code. The matrix elements of the symmetry generators can be deduced from the fusion rules in Table.\ref{table: S4-S3 fusion} similar to the previous examples.

\subsubsection{$H=S_3,~ K=\bZ_4$}
Here the symmetry boundary is as before, $\e{C}=\fk{C}[S_4,S_3,0,0]$. We take $K=\bZ_4=\<(1,2,3,4)\>$. Since $H\cap K=\{e\}$ and $HK=S_4$. From our discussion in Sec.~\ref{subsec: minimal}, we see this this choice leads to an SPT of $\fk{C}[S_4,S_3,0,0]$. 

In fact there are four SPTs for this choice, since the physical boundary can be twisted by a class $\psi\in \cc{H}^3[\bZ_4,\bC^\times]$.

\subsubsection{$H=S_3,~K=\bZ_2\times\bZ_2$}

We take $K=\bZ_2\times\bZ_2=\<(1,2),(3,4)\>$. Again we have $H\cap K=\{e\}$ and $HK=S_4$. This define another family of SPTs of  $\fk{C}[S_4,S_3,0,0]$.

There are eight SPTs, since the physical boundary can be twisted by a class $\psi\in \cc{H}^3[\bZ_2\times \bZ_2,\bC^\times]\simeq \bZ_2^3$.

\begin{remark}
    The above two families of SPTs are not gauge related to any single family of invertible SPTs. More precisely, if we change the symmetry boundary to be $2\vc_{S_4}$, then these two families are mapped to gapped phases with different unbroken symmetry groups, one with $\bZ_4$, the other with $\bZ_2\times \bZ_2$. Therefore these two families of SPTs can not be obtained by gauging a single family of invertible SPTs. It would be particularly interesting to explore the gapless edge modes between them, which should be distinct from the edge modes of any 2+1D invertible SPTs.
\end{remark}
\section{Topological holography for 2+1-D gapless phases\label{sec: gapless}}
We now turn to the study of topological holography for 2+1D gapless phases. A complete mathematical theory for gapless phases and their interplay with generalized symmetries remains an open challenge. Within the scope of this paper, we show that the \symto club-sandwich construction determines the \textit{topological skeleton} of gapless phases. 

The concept of topological skeleton was introduced in a series of papers~\cite{Kong_2022_cql1,Kong_2024_cql2}. The spirit of ~\cite{Kong_2022_cql1,Kong_2024_cql2} is that all the physical observables of a gapless phase can be divided into two pieces:
\begin{itemize}
    \item The information of local observables such as the
OPE of local fields. This is called the \textit{local quantum symmetry} of the gapless phase. In 1+1D, this can be the chiral algebra or full field algebra of a CFT.
\item The information of all the topological defects. This is called the \textit{topological skeleton} of the gapless phase. 
\end{itemize}

It is expected that the topological skeleton of a gapless phase  has a higher categorical structure. These two pieces of information are not independent and must be compatible in certain ways. Studies on 1+1D phases with symmetry have shown that the topological skeleton of a phase(gapped or gapless) is intimately related to condensable algebras in the \symto~\cite{wen2023classification11dgaplesssymmetry,Kong_2022_enriched}. This is best illustrated through the club-sandwich construction~\footnote{In~\cite{Kong_2022_cql1,Kong_2024_cql2,Kong_2022_enriched}, this is called the ``topological wick rotation" principle.}. 

\subsection{The club-sandwich construction}
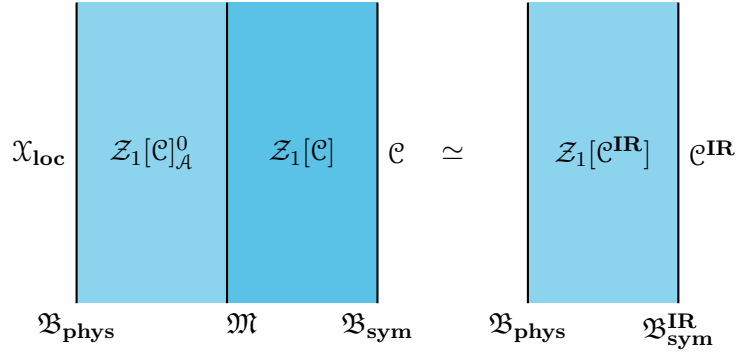
\begin{figure}
    \centering
        \begin{tikzpicture}[baseline=(current  bounding  box.center)]
\fill[zima!70] (0,0) rectangle (2,4);
\fill[zima] (2,0) rectangle (4,4);
\draw[ thick]  (0,0) -- (0,4);
\draw[ thick] (4,0) -- (4,4);
\draw[thick]  (2,0) -- (2,4);
\node[label, below] at (0,0){$\fk{B}_{\mathbf{phys}}$};
\node[label,left] at (0,2){$\e{X}_{\mathbf{loc}}$};
\node[label, below] at (2.2,0){$\fk{M}$};
\node[label, below] at (4,0){$\fk{B}_\sym$};
\node[label] at (1,2){$\Z_1[\e{C}]_{\e{A}}^0$};
\node[label] at (3,2){$\Z_1[\e{C}]$};
\node[label, right] at (4,2){$\e{C}$};
\node[label] at (5,2){$\simeq$};
\fill[zima!70] (6,0) rectangle (8,4);
\draw[thick] (6,0) -- (6,4);
\draw[thick] (8,0) -- (8,4);
\node[label, below] at (6,0){$\fk{B}_\phys$};
\node[label, below] at (8,0){$\fk{B}^{\mathbf{IR}}_{\sym}$};
\node[label] at (7,2){$\Z_1[\e{C}^{\IR}]$};
\node[label, right] at (8,2){$\e{C}^\IR$};
\end{tikzpicture}
    \caption{The club-sandwich construction separates the topological skeleton from the local quantum symmetry of a gapless phase. The left boundary encodes all the local data of the gapless phase such as OPE of local fields. The domain wall $\fk{M}$ encodes the topological skeleton of the gapless phase, including the topological sectors of states and operators. By fusing the domain wall $\fk{M}$ with the symmetry boundary $\fk{B}_{\sym}$, we obtain an ordinary sandwich that describes the IR symmetry of the gapless phase. In particular, the new symmetry boundary describes the IR symmetry of the gapless phase, denoted as $\fk{B}_{\sym}^\IR$.}
    \label{fig: sandwich_4}
\end{figure}

The basic setup is summarized in Fig.~\ref{fig: sandwich_4}. The bulk is divided into regions, separated by a gapped domain wall $\fk{M}$. To the right of this wall hosts the \symto $\Z_1[\e{C}]$ with its right boundary being the symmetry boundary $\e{C}$. The left region is obtained through a condensable algebra $\e{A}$ in $\Z_1[\e{C}]$ and is denoted as $\Z_1[\e{C}]_{\e{A}}^0$. The left boundary of the left region is a gapless boundary of $\Z_1[\e{C}]_{\e{A}}^0$. This gapless boundary should be irreducible in some sense: all operators of $\Z_1[\e{C}]_{\e{A}}^0$ should act nontrivially on this gapless boundary, otherwise we can further reduce $\Z_1[\e{C}]_{\e{A}}^0$ by a condensation.

This construction describes the topological skeleton of a 2+1D phase as follows~\cite{Kong_2022_cql1,Kong_2024_cql2,Kong_2022_enriched}.
\begin{itemize}
    \item The \symto $\Z_1[\e{C}]$ describes the topological sectors of operators, including both local and non-local operators symmetric under $\e{C}$. 
    \item The domain wall $\fk{M}$ describes the topological sectors of states of the 2+1D phase. A topological sector of state is a subspace of the Hilbert space that is invariant under the action of local symmetric operators.
    \item The operators from $\Z_1[\e{C}]$ can move to the domain wall and act on objects on the wall. This corresponds to changing the topological sector of states by acting with topological sector of operators.
    \item The operators from $\Z_1[\e{C}]$ that act trivially on the wall $\fk{M}$ are the topological sectors of operators that leave the vacuum sector invariant. These operators are therefore condensed on the wall.
\end{itemize}

If we shrink the $\Z_1[\e{C}]$ region and fuse the domain wall $\fk{M}$ with the symmetry boundary $\fk{B}_{\sym}$, we obtain an effective ordinary sandwich. See Fig.~\ref{fig: sandwich_4}. The new  symmetry boundary describes the effective symmetry of the 2+1-D phase that acts nontrivially in the IR, called the IR symmetry, denoted by $\e{C}^\IR$. It is given by a relative tensor product
\begin{align}
    \e{C}^\IR=\fk{M}\boxtimes_{\Z_1[\e{C}]} \e{C}.
\end{align}

If we consider a topological defect on the original symmetry boundary $\e{C}$, then after shrinking the $\Z_1[\e{C}]$ region it will become a topological defect on the new symmetry boundary $\e{C}^\IR$. This process defines a functor 
\begin{align}
    \e{R}: \e{C}\to \e{C}^\IR,
\end{align}
which describes how the exact symmetry $\e{C}$ reduces to the IR symmetry $\e{C}^\IR$ as we zoom out. 
\section{Higher condensation theory for 3+1D topological orders\label{sec: higher condensation}}
The crucial input in the club-sandwich construction is a condensable algebra in the \symto. Therefore the theory of condensation is critical to the topological holography paradigm for gapless phases.  Here we review the part of this elegant theory~\cite{kong2025highercondensationtheory} relevant for this work, namely $E_2$-condensation in 3+1D topological orders. More details can be found in Appendix~\ref{app: condensation}.

At the microscopic level, a condensation of topological defect amounts to turning on some commuting projector Hamiltonian to trivialize certain topological defects. At the macroscopic level, this process is described by algebras in the category of topological defects~\cite{gaiotto2019condensationshighercategories,kong2025highercondensationtheory}. In 3+1D there are point-like as well as string-like defects,  therefore we naturally have two types of condensable algebras in 3+1D. To describe condensation of strings, one considers the so-called condensable $E_2$-algebras~\footnote{A particle condensation is described by a so-called $E_3$-condensable algebra. Particle condensation can always be reduced to string condensation, see discussion near the end of this section.}. The collection of all condensable $E_2$-algebras in a 3+1D topological order $\e{C}$ will be denoted as $\alg^c_{E_2}(\e{C})$.

After condensing an algebra $\e{A}\in\alg^c_{E_2}(\e{C})$, we have the following physical picture.
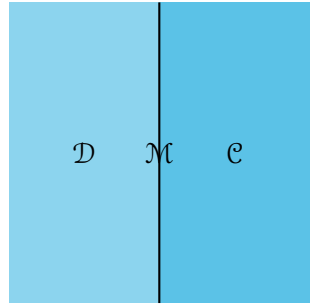
\begin{figure}[h]
    \centering
        \begin{tikzpicture}[baseline=(current  bounding  box.center)]

\fill[zima!70] (0,0) rectangle (2,4);
\fill[zima!70] (0,0) rectangle (2,4);
\fill[zima] (2,0) rectangle (4,4);
        \draw[thick] (2,0) -- (2,4);
\node[label] at (2,2){$\e{M}$};
\node[label] at (1,2){$\e{D}$};
\node[label] at (3,2){$\e{C}$};
\end{tikzpicture}
    \label{fig: condensation}
\end{figure}

The objects of interests are the condensed phase $\e{D}$  and the domain wall $\e{M}$ between the original phase and the condensed phase. These objects are described by category of modules over $\e{A}$.
\begin{ptheorem}
    Let $\e{C}$ be a 3+1D topological order, i.e. a non-degenerate braided fusion 2-category. Then a string condensation is specified by a condensable $E_2$-algebra $\e{A}\in\alg^c_{E_2}(\e{C})$, and we have:
    \begin{itemize}
        \item The condensed phase is a new topological order $\e{D}$, described by the braided fusion 2-category of local modules over $\e{A}$: $\e{D}=\Mod^0_{\e{A}}(\e{C})$.
        \item The domain wall between the old phase and the new phase is described by the fusion 2-category of modules over $\e{A}$: $\e{M}=\Mod_{\e{A}}(\e{C})$.
        \item Moving defects in $\e{C}$ onto the wall defines a functor from $\e{C}$ to $\e{M}$, given by 
        \begin{align}
            X\mapsto X\otimes \e{A}
        \end{align}
        \item Conversely, any defect on the wall $\e{M}$ can be moved into $\e{C}$, given by the forgetful functor 
        \begin{align}
            F:\Mod_{\e{A}}(\e{C})\to \e{C}\label{eq: forget}
        \end{align}
        \item Defects in $\e{D}$ can move onto the wall by the inclusion $\Mod^0_{\e{A}}(\e{C})\to \Mod_{\e{A}}(\e{C})$, and then can move freely into $\e{C}$ via the forgetful functor~\eqref{eq: forget}.
        \item By folding, $\e{M}$ is a gapped boundary of $\e{C}\boxtimes \e{D}^{op}$. By the bulk-boundary relation of topological orders we have
        \begin{align}
            \Z_1[\Mod_{\e{A}}(\e{C})]\simeq \e{C}\boxtimes \Mod^0_{\e{A}}(\e{C})^{op}
        \end{align}
    \end{itemize}
\end{ptheorem}
A special case of string condensation is when the condensed phase is trivial, the corresponding condensable algebra in this case is called Lagrangian.
\begin{definition}
    A condensable $E_2$-algebra is called Lagrangian if $\Mod^0_{\e{A}}(\e{C})=2\vc$.
\end{definition}
For a Lagrangian condensation, the domain wall $\e{M}$ becomes a gapped boundary of $\e{C}$. The bulk-boundary relation of topological order states all gapped boundaries arise in this way. 
\begin{ptheorem}
    Let $\e{M}$ be a gapped boundary of $\e{C}$, then there is a Lagrangian algebra $\e{A}$ in $\e{C}$ such that 
    \begin{align}
        \e{M}=\Mod_{\e{A}}(\e{C}),
    \end{align}
    and we have 
    \begin{align}
        \Z_1[\Mod_\e{A}(\e{C})]\simeq \e{C}.
    \end{align}
\end{ptheorem}

\paragraph*{\textbf{Sequential condensation}} Consider starting with a 3+1D topological order $\e{C}$, perform a string condensation $\e{A}_1$ to obtain phase $\e{D}$, and then perform another string condensation $\e{A}_2$ in $\e{D}$ to obtain phase $\e{E}$. This two-step condensation is in fact equivalent to a single condensation in $\e{C}$, described as follows.
\begin{definition}
    Let $\e{A}\in\alg^c_{E_2}(\e{C})$, a condensable $E_2$-algebra \textit{over} $\e{A}$ is a condensable algebra $\e{B}$ together with a map between condensable $E_2$-algebras: $f: \e{A}\to \e{B}$.
\end{definition}
\begin{theorem}\label{thm: seq_cond}
    Let $\e{A}\in\alg^c_{E_2}(\e{C})$, then a condensable algebra $\e{B}$ in $\Mod^0_{\e{A}}(\e{C})$ is the same as a condensable algebra in $\e{C}$ over $\e{A}$, and we have 
    \begin{align}
        \Mod^0_{\e{B}}(\Mod^0_{\e{A}}(\e{C}))\simeq \Mod^0_{\e{B}}(\e{C}).
    \end{align}
\end{theorem}
All the above statements are natural generalization of the anyon condensation theory to 3+1D. However, the structure of condensable algebras is more complicated in 3+1D, thus it is desirable to have systematical methods for constructing condensable algebras, as well as for computing the condensed phase $\Mod^0_{\e{A}}(\e{C}))$. In the next section, we focus on the case $\e{C}=\Z_1[2\vc_G]$, for which more explicit description of condensable $E_2$-algebras can be given. We also describe how to compute the condensed phase for a general string condensation in $\Z_1[2\vc_G]$. 
\begin{remark}
    We briefly comment on particle condensation in 3+1D here. For a 3+1D topological order $\e{C}$, particles live in $\Omega\e{C}$,  a symmetric fusion category. A particle condensation is defined by a condensable $E_3$-algebra in $\Omega\e{C}$, $\e{A}\in\alg^c_{E_3}(\Omega\e{C})$. A particle condensation can be treated as a string condensation: we can first condense the particles along one direction, obtaining the so-called condensation descendant of the particles. Then condensing particles is the same as condensing this descendant string. Mathematically, the descendant string is $\Sigma\e{A}\in \alg^c_{E_2}(\e{C})$. For example, the Cheshire string $\bbm{1}_c$ in the 3+1D toric code is the condensation descendant of the gauge charge: $\bbm{1}_c=\Sigma(1+e)$. Condensing the particle $e$ is then the same as condensing the string $\bbm{1}_c$.
\end{remark}
\section{String condensation in $\Z_1[2\vc_G]$\label{sec: string cond}}
 In this section we review the structure of condensable algebras in $\Z_1[2\vc_G]$~\cite{xu2024etalealgebrasbosonicfusion}. We also provide a systematic method for determining the condensed phase for any condensable algebra in $\Z_1[2\vc_G]$. 
\subsection{Condensable $E_2$-algebras in $\Z_1[2\vc_G]$}
The full structure of the braided fusion 2-category $\Z_1[2\vc_G]$ can be found in ~\cite{Kong_2020_center,wen2025stringcondensationtopologicalholography}. We note that it can be formulated as the 2-category of $G$-crossed finite semi-simple 1-categories. From this formulation, it follows from the definition of condensable $E_2$-algebras Def.~\ref{def: E_2}, Def~\ref{def: c_E_2} that we have the following fact.
\begin{fact}
    A condensable $E_2$-algebra in $\Z_1[2\vc_G]$ is a $G$-crossed braided multi-fusion category such that
     \begin{align*}
       \text{ (C1) ~The $G$-action permutes the simple summands of the monoidal unit transitively.}
    \end{align*}
\end{fact}
This follows directly from the definition of condensable $E_2$-algebras and the description of the braided fusion 2-category $\Z_1[2\vc_G]$. Roughly speaking, the underlying object of the algebra is by definition a $G$-crossed finite semi-simple category, then the algebra structure(product, unit, unitors, associator) equips it with a $G$-crossed monoidal structure. The braiding in Def.~\ref{def: E_2} further equips it with a $G$-crossed braiding. Demanding the algebra to be separable amounts to demanding it to be multi-fusion. Finally, condition (C1) is equivalent to the connectivity condition in Def.~\ref{def: c_E_2}. 

    A physical interpretation of the multi-fusion structure of the condensable algebra is as follows. If we take the \symto sandwich construction and put a condensable algebra at the physical boundary, and fix the symmetry boundary to be $2\vc_G$, then we obtain a 2+1-D phase with a global symmetry $G$. This phase breaks the symmetry $G$ exactly when the condensable algebra at the physical boundary is multi-fusion, in which case the simple summands of the monoidal unit correspond to the vacuua of the phase, and they are permuted by the $G$-action. This $G$-set of simple summands must be irreducible and is equal to the coset $G/H$ for some subgroup $H<G$, which is the unbroken subgroup of the 2+1-D phase.

Let $\e{A}$ be a condensable algebra in $\Z_1[2\vc_G]$, the full sub-category generated by the simple summands of the monoidal unit of $\e{A}$ must be equivalent to a functor category $\mathbf{Func}[G/H,\vc]$ for some subgroup $H<G$. This sub-category is by itself a $G$-crossed braided multi-fusion category satisfying (C1), i.e. a condensable algebra in $\Z_1[2\vc_G]$. Thus we have inclusion of condensable algebras $\mathbf{Func}[G/H,\vc]\to \e{A}$. By the sequential condensation theorem~\ref{thm: seq_cond}, condensing $\e{A}$ can then be divided into two steps, where in the first step we condense the sub-algebra $\mathbf{Func}[G/H,\vc]$. This step is rather simple(proof can be found in, e.g.~\cite{kong2025highercondensationtheory}):
 \begin{align}
        \Mod^0_{\mathbf{Func}[G/H,\vc]}(\Z_1[2\vc_G])\simeq \Z_1[2\vc_H^{}].
\end{align}

The second step is then to  viewed $\e{A}$ as a condensable algebra in $\Z_1[2\vc_H^{}]$. By results in~\cite{xu2024etalealgebrasbosonicfusion}, viewed as a condensable algebra in $\Z_1[2\vc_H^{}]$, $\e{A}$ is now a $H$-crossed braided \textit{fusion} category. Thus we have the following equivalent characterization of condensable algebras in $\Z_1[2\vc_G^\pi]$.
\begin{fact}
    Condensable $E_2$-algebras in $\Z_1[2\vc_G]$ are labeled by pairs $(H,\e{B})$, where $H<G$ is a subgroup, and $\e{B}$ is a $H$-crossed braided fusion category.
\end{fact}
We say the condensable algebra is \textit{strongly connected} if $H=G$ in the above description. In this case, it is described by a $G$-crossed braided \textit{fusion} category. We also stress here that here the $H$-crossed braided fusion category $\e{B}$ does not need to have faithful grading. In the decomposition
\begin{align}
    \e{B}=\oplus_h \e{B}_h,
\end{align}

we do not demand that all components are non-zero. Also, the trivial component $\e{B}_e$, being a braided fusion category, does not have to be non-degenerate. In fact, these two conditions(faithful grading and non-degenerate trivial component) are met exactly when the condensable algebra is Lagrangian~\cite{Wen_BIMSA,xu2024etalealgebrasbosonicfusion}. 

\begin{theorem}\label{thm: Lagrangian}
    Let $(H,\cc{B})\in\alg^c_{E_2}(\Z_1[2\vc_G^\pi])$ be described above. Then it is Lagrangian iff the grading of $\cc{B}$ is faithful and $\cc{B}_e$ is non-degenerate.
\end{theorem}
\subsection{Computing the condensed phase}
For a non-Lagrangian condensable algebra, we are  interested in the condensed phase. In the \symto construction, the condensed phase describes the low energy symmetry of the corresponding gapless phase. When the \symto is $\Z_1[2\vc_G]$, there is a rather complete answer.

There is a standard Morita equivalence: $2\vc_G\sim \Sigma\rep(G)$, where $\Sigma\rep(G)$ is the fusion 2-category of $\rep(G)$-module categories. This leads to an equivalence between their Drinfeld centers:
\begin{align}
  \Phi:  \Z_1[2\vc_G]\simeq \Z_1[\Sigma\rep(G)]
\end{align}

Consequentially there is an equivalence between condensation in $\Z_1[2\vc_G]$ and condensation in $\Z_1[\Sigma\rep(G)]$. The description of condensable $E_2$-algebras in $\Z_1[\Sigma\rep(G)]$ and their condensed phase are rather simple~\cite{D_coppet_2024}:
\begin{fact}
    A condensable $E_2$-algebra in $\Z_1[\Sigma\rep(G)]$ is a braided fusion category $\cc{B}$ together with a braided functor from $\rep(G)$.  The condensed phase is 
\begin{align}
    \Mod^0_{\cc{B}}(\Z_1[\Sigma\rep(G)])\simeq \Z_1[\Sigma\cc{B}].
\end{align}
\end{fact}
Let $\e{A}$ be a strongly-connected condensable algebra in $\Z_1[2\vc_G]$, i.e. a $G$-crossed braided fusion category. Under the equivalence $\Phi$, it is mapped to a condensable algebra in $\Z_1[\Sigma\rep(G)]$, i.e. a braided fusion category with a braided functor from $\rep(G)$. This is given by the equivariantization of $\e{A}$, $\Phi(\e{A})=\e{A}^G$~\cite{D_coppet_2024}. Therefore we have 
\begin{align}
    \Mod^0_{\e{A}}(\Z_1[2\vc_G])\simeq \Z_1[\Sigma\e{A}^G].\label{eq: condensed_1}
\end{align}

Eq.~\eqref{eq: condensed_1} provides an answer to the condensation problem. However, it is not an ideal answer. For instance, in many cases the Müger center $ \Z_2\e{A}^G$ is Tannakian, then we know that $\Z_1[\Sigma\e{A}^G]$ must take the form $\Z_1[2\vc_K^\omega]$ for some $K$ and some $\omega\in \cc{H}^4[K,U(1)]$, and we would want to determine $K$ and $\omega$ from the condensable algebra $\e{A}$. In other words, if we know that the condensed phase is a Dijkgraaf-Witten theory, we wish to determine the gauge group and the twist. Using the classification of fusion 2-categories developed in~\cite{décoppet2024classificationfusion2categories}, this can be done as follows. 

\begin{lemma}
    Let $\cc{B}$ be a braided fusion category with a Tannakian Müger center: $\Z_2\cc{B}=\rep(K)$. Then the de-equivariantization $\cc{B}_K$ is a non-degenerate braided fusion category carrying a natural $K$-action. Let $\omega\in \cc{H}^4[K,U(1)]$ be the $\cc{O}_4$-obstruction of this $K$-action~\footnote{See appendix~\ref{app: extension theory} for the definition of the $\cc{O}_4$-obstruction.}. Then there is an equivalence
    \begin{align}
        \Z_1[\Sigma\cc{B}]\simeq \Z_1[2\vc_K^\omega]
    \end{align}
\end{lemma}

Therefore, we obtain the following practical algorithm for computing the condensed phase.
\paragraph*{\textbf{Determining the condensed phase.}}
Let $\e{A}\in \alg^c_{E_2}(\Z_1[2\vc_G])$ be a condensable algebra such that $\Z_2\e{A}_e$ is Tannakian. Then the condensed phase is a finite gauge theory,
\begin{align}
    \Mod^0_{\e{A}}(\Z_1[2\vc_G])=\Z_1[2\vc_K^\omega],\label{eq: condensed_phase_1}
\end{align}
where $K$ is determined from $\Z_2(\e{A}^G)=\rep(K)$, and $\omega\in \cc{H}^4[K,U(1)]$ is the $\cc{O}_4-$obstruction of the $K$-action on $(\e{A}^G)_K$. Some useful results about computing $\cc{O}_4-$obstruction of braided fusion category with group action can be found in~\cite{Cui_2016}.
\paragraph*{\textbf{The case $\e{A}_e$ is non-degenerate.}}
In the examples studied in this work, the trivial component of the condensable algebra $\e{A}_e$ will always be non-degenerate. In this case we may give a more concrete description for the condensed phase. 

Let $\e{A}\in \alg^c_{E_2}(\Z_1[2\vc_G])$ be a strongly connected condensable algebra such that $\e{A}_e$ is non-degenerate. Then the support of the $\e{A}$ must be a normal subgroup $N\lhd G$. By results in~\cite{drinfeld2010braidedfusioncategoriesi}, we have $\Z_2(\e{A}^G)=\rep(G/N)$. Therefore in this case we have 
\begin{align}
     \Mod^0_{\e{A}}(\Z_1[2\vc_G])=\Z_1[2\vc_{G/N}^\omega]\label{eq: condensed_phase_2}
\end{align}
where $\omega$ is the $\cc{O}_4-$anomaly of the $G/N$-action on $(\e{A}^G)_{G/N}\simeq \e{A}^N$. In other words,  if the support of $\e{A}$ is $N$, then we may view it as an $N$-crossed braided fusion category, and equivariantize with respect to $N$. The result $\e{A}^N$ now carries an action by $G/N$, and its $\cc{O}_4$-obstruction is exactly the twist of the condensed phase.

\section{Minimal condensable algebras: Cohomology method\label{sec: minimal_cd_alg}}
\hide{In order to construct 2+1-D gapless phases from the \symto club-sandwich construction, we would want a systematic method for constructing condensable algebras in the \symto. A even more ambitious goal is to classify all condensable algebras in the \symto. However in general this is very difficult, since there are always infinite many condensable algebras in any 3+1D topological order. This difficulty is essentially due to the presence of infinite many 2+1-D topological orders. However, if we focus on those condensable algebras that are dual to 2+1-D phases with no topological orders, this goal becomes reachable. A similar situation occurs for gapped $G$-symmetric phases in 2+1-D, where if we restrict to those without topological order, i.e. SPTs, then a finite, algebraic classification becomes possible. }
From the general principle of topological holography, we expect a non-Lagrangian condensable algebra in the \symto to describe the topological skeleton of a gapless 2+1-D phase with symmetry. However, not all such  condensable algebras should be considered ``gapless SPT". Similar to the gapped case, we should expect there to be a notion of ``gapless symmetry enriched topological order"(gSET) in 2+1-D, and gSPT should be a special kind of gSET, where the gapped sector has trivial topological order. 

In this section, we introduce the class of condensable algebras in $\Z_1[2\vc_G]$ that we expect to correspond to 2+1-D $G$-SPTs/gSPTs based on the picture of club-sandwich. We call them minimal condensable algebras. The definition is:
\begin{definition}
   Recall a condensable $E_2$-algebra in $\Z_1[2\vc_G]$ is labeled as $(H,\cc{B})$, where $\cc{B}$ is an  $H$-crossed braided fusion category. We say the condensable $E_2$-algebra is minimal if $\cc{B}_e=\vc$.
\end{definition}
In the club-sandwich picture Fig.~\ref{fig: sandwich_4}, the braided fusion category $\cc{B}_e$ describes the intrinsic 2+1-D topological order on the domain wall $\fk{M}$. If $\fk{M}$ has nontrivial topological order, then the gapped sector of the corresponding gapless phase has nontrivial topological order and we consider such gapless phase as ``gapless symmetry-enriched topological order"(gSET).

One of the main results of this paper is an exact duality between minimal condensable algebras and 2+1-D $G$-gSPTs. To establish this correspondence, it is crucial to understand the mathematical structure of minimal condensable algebras, which is the goal of this section. We give a cohomological characterization and classification of minimal condensable algebras using the theory of crossed extension of braided fusion categories and the LHS spectral sequence. We also provide a method for computing the condensed phase of any minimal condensable algebra via spectral sequence.

\subsection{Structure of minimal condensable algebras}
We only need to focus on the case $H=G$ , the general case can be dealt with using sequential condensation theorem~\ref{thm: seq_cond}. In this case, the condensable algebra is a $G$-crossed braided fusion category $\cc{B}$ such that the trivial component is trivial $\cc{B}_e=\vc$. Equivalently $\cc{B}$ is a $G$-crossed extension of $\vc$. It is well-known that if the grading of the $G$-crossed extension is faithful, then such extensions are classified by the group cohomology $\cc{H}^3[G,\bC^\times]$. However we see from Theorem~\ref{thm: Lagrangian} that this is only true when the condensable algebra is Lagrangian. For a general condensable algebra, the grading of $\cc{B}$ is supported on a normal subgroup $N\lhd G$.

We denote the set of equivalent classes of minimal condensable algebras supported on $N$ as $G\mathbf{CrExt}(\vc,N)$, which stands for ``$G$-crossed extensions of $\vc$ supported on $N$". We start with the following observation.
\begin{lemma}
    A $G$-crossed braided fusion category $\cc{B}$ with non-degenerate $\cc{B}_e$  and support $N\lhd G$ is equivalent to its trivial component $\cc{B}_e$ plus the following data:
    \begin{enumerate}
        \item A faithful $N$-crossed extension $\cc{B}_e^\times$.
        \item A $G/N$ action on $(\cc{B}_e^\times)^N$:
        \begin{align}
            \Phi: G/N\to  \underline{\Aut_{br}}((\cc{B}_e^\times)^N)
        \end{align}
        such that $G/N$ acts on $\rep(N)\subset (\cc{B}_e^\times)^N$ as the canonical action induced by the action of $G/N$ on $N$.
    \end{enumerate}
    \proof
    For every $G$-crossed braided fusion category $\cc{B}$ supported on $N\lhd G$, by restricting to $N$-action we an view $\cc{B}$ as an $N$-crossed braided fusion category with faithful grading.  It is standard that if $\cc{B}$ has a $G$-action, then $\cc{B}^N$ has a natural $G/N$-action that acts on $\rep(N)$ canonically~\cite{drinfeld2010braidedfusioncategoriesi,Cui_2016}.

    Conversely, if the trivial component $\cc{B}_e$ and the above data are given, the $G$-crossed braided fusion supported on $N$ can be recovered as follows. Using the $G/N$-action on $(\cc{B}_e^\times)^N$, we may form the equivariantization $((\cc{B}_e^\times)^N)^{G/N}$. Since $\rep(N)\subset (\cc{B}_e^\times)^N$, we have 
    \begin{align}
        \rep(N)^{G/N}\simeq \rep(G)\subset ((\cc{B}_e^\times)^N)^{G/N}.
    \end{align}
    
    Thus $((\cc{B}_e^\times)^N)^{G/N}$ contains $\rep(G)$. We may then de-equivariantize with respect to $G$ to obtain $(((\cc{B}_e^\times)^N)^{G/N})_G$. This is a $G$-crossed braided fusion category. Since $\Z_2(((\cc{B}_e^\times)^N)^{G/N})=\rep(G/N)$, the support of $(((\cc{B}_e^\times)^N)^{G/N})_G$ is exactly $N$.\qed
\end{lemma}
Specializing to the case $\cc{B}_e=\vc$, a faithful $N$-crossed extension is determined by a class $\omega\in \cc{H}^3[N,\bC^\times]$, and the corresponding $N$-crossed extension is $\vc_N^\omega$. The $N$-equivariantization of $\vc_N^\omega$ is know: $(\vc_N^\omega)^N\simeq \Z_1[\vc_N^\omega]$.  We obtain the following characterization of minimal condensable algebras in $\Z_1[2\vc_G]$.
\begin{corollary}\label{coro: GCrExt}
    An element $\cc{B}\in G\mathbf{CrExt}(\vc,N)$ is determined by 
    \begin{enumerate}
        \item A class $\omega\in\cc{H}^3[N,\bC^\times]$.
        \item A homomorphism of 2-groups 
        \begin{align}
           \underline{\Phi}: G/N \to \underline{\Aut_{br}}(\Z_1[\vc_N^\omega])\label{eq: 2-group_map}
        \end{align}
          such that $G/N$ acts on $\rep(N)\subset \Z_1[\vc_N^\omega]$ canonically.
    \end{enumerate}
\end{corollary}
\subsection{The LHS spectral sequence\label{sec: LHS}}
We now explain how the above characterization of $G\mathbf{CrExt}(\vc,N)$ may be reformulated in terms of the LHS spectral sequence. We have $\underline{\Aut_{br}}(\Z_1[\vc_N^\omega])\simeq \underline{\mathbf{BrPic}}(\vc_N^\omega)$. Therefore the 2-group morphism $\underline{\Phi}$ can be viewed as
\begin{align}
    \underline{\Phi}: G/N\to \underline{\mathbf{BrPic}}(\vc_N^\omega).
\end{align}

We understand that a morphism of 3-groups 
\begin{align}
    \underline{\underline{\Phi}}: G/N\to \underline{\underline{\mathbf{BrPic}}}(\vc_N^\omega)
\end{align}
corresponds to a $G/N$-extension of $\vc_N^\omega$. As explained in~\cite{etingof2009fusioncategorieshomotopytheory}, such an extension amounts to a group extension of $G/N$ by $N$ together with an extension of $\omega$ to a cocycle in $\cc{H}^3[G,\bC^\times]$. In our case, the group extension is fixed to be $1\to N\to G\to G/N\to1$. Therefore the rest of information in the 3-group morphism is just an extension of $\omega$ from a $N$-cocyle to a $G$-cocycle. From the theory of spectral sequence, we know that such an extension exists if and only if $d_2\omega=0$ in $E_2^{2,2}$, $d_3\omega=0$ in $E_3^{3,1}$, and $d_4\omega=0$ in $E_4^{4,0}$. When these conditions are met, an extension of $\omega$ may be constructed step-by-step as follows. In the first step, we examine $d_2\omega\in H^2[G/N,H^2[N,\bC^\times]]$. The condition $d_2\omega=0$ means certain equation involving $\omega$ has a solution. We need to choose a solution $\gamma$, then $\omega,\gamma$ together defines a class $i^{\gamma,\omega}\in E_2^{3,1}=H^3[G/N,H^1[N,\bC^\times]]$. This element $i^{\gamma,\omega}$ is in the kernel of $d_2$, and its image in $E_3^{3,1}$  is $d_3\omega$. The condition $d_3\omega=0$ means that we can choose $\gamma$ so that $i^{\gamma,\omega}=0$ in $E_2^{3,1}$. We make such a choice. Then $i^{\gamma,\omega}=0$ means certain equation involving $\omega,\gamma$ has solution and we choose a solution $p$. From $\omega,\gamma,p$, we can construct a class $j^{\omega,\gamma,p}\in E^{4,0}_2=H^4[G,\bC^\times]$. The class $j^{\omega,\gamma,p}$ is in the kernel of $d_2,d_3$, and its image in $E_4^{4,0}$ would be $d_4\omega$. The condition $d_4\omega=0$ means that for certain choice of $\gamma,p$ we can make $j^{\omega,\gamma,p}=0$ in $E_2^{4,0}$. We make such a choice. Then $j^{\omega,\gamma,p}=0$ means certain equation has solution, and we choose a solution $\beta$. Together the triple $(\gamma,p,\beta)$ gives a desired extension of $\omega$ to $G$. Cochain level expression of the classes $(\gamma,p,\beta)$ and the equations they satisfy may be found in~\cite{wang2021domainwalldecorationsanomalies}.

The obstruction classes $d_3\omega,d_4\omega$ correspond to the obstruction of lifting $\Phi$ to a 2-group homomorphism and to a 3-group homomorphism, often called the $\cc{O}_3$-obstruction and the $\cc{O}_4$-obstruction.
In the description of $G\mathbf{CrExt}(\vc,N)$, Corollary~\ref{coro: GCrExt}, $\underline{\Phi}$ is only a 2-group homomorphism, and the obstruction of lifting it to a 3-group homomorphism is exactly the twist of the condensed phase by Eq.~\eqref{eq: condensed_phase_2}. Therefore we have the following.
\begin{theorem}
    Elements in $G\mathbf{CrExt}(\vc,N)$ are parameterized by $\omega\in \cc{H}^3[N,\bC^\times]$ such that $d_2\omega=0,d_3\omega=0$ together with a solution to these equations $(\gamma,p)$. Denote the corresponding condensable algebra as $\vc_N^{(\omega,\gamma,p)}$, the condensed phase is
     \begin{align}
\Mod_{\vc_N^{(\omega,\gamma,p)}}^0(\Z_1[2\vc_G])\simeq \Z_1[2\vc_{G/N}^{d_4\omega}].
    \end{align}
    Here by $d_4\omega$ we mean a representative of it in $\cc{H}^4[G/N,\bC^\times]$ computed using the solution $(\gamma,p)$.
\end{theorem}
\subsection{Group structure of minimal condensable algebras and an exact sequence}
The set of equivalent classes of minimal condensable algebras with support $N$, $G\mathbf{CrExt}(\vc,N)$, actually has a natural group structure. Consider two elements in $G\mathbf{CrExt}(\vc,N)$, $\vc_N^{(\omega,\gamma,p)},\vc_N^{(\omega',\gamma',p')}$. Take the diagonal components of $\vc_N^{(\omega,\gamma,p)}\boxtimes \vc_N^{(\omega',\gamma',p')}$, we obtain another element of $G\mathbf{CrExt}(\vc,N)$. This defines a product $\circ$ on $G\mathbf{CrExt}(\vc,N)$, which is clearly associative and commutative. In terms of the data $\omega,\gamma,p$ we simply have $\vc_N^{(\omega,\gamma,p)}\circ \vc_N^{(\omega',\gamma',p')}=\vc_N^{(\omega+\omega,\gamma+\gamma',p+p')}$. Therefore $G\mathbf{CrExt}(\vc,N)$ is naturally an abelian group. We prove an exact sequence relating $G\mathbf{CrExt}(\vc,N)$ to ordinary group cohomology, which enables us to compute $G\mathbf{CrExt}(\vc,N)$ as a set.

There is a natural group homomorphism from $G\mathbf{CrExt}(\vc,N)$ to $\cc{H}^4[G/N,\bC^\times]$ given by taking the obstruction class of the 2-group morphism $\underline{\Phi}$. This obstruction class is exactly the obstruction $j^{\omega,\gamma,p}$ of lifting $\omega$ to $G$. Since $j^{\omega,\gamma,p}$ is linear in $\omega,\gamma,p$, this defines a group homomorphism
\begin{align}
    \cc{O}_4: G\mathbf{CrExt}(\vc,N)\to \cc{H}^4[G/N,\bC^\times].
\end{align}

We claim the image of this map exactly contain those elements that trivialize in $\cc{H}^4[G,\bC^\times]$. Indeed, for any element in $G\mathbf{CrExt}(\vc,N)$, the 2-group morphism $\underline{\Phi}$ gives us ``almost an extension of $\vc_N^\omega$": $\vc_G^\alpha$, except that the associator $\alpha_{g,h,k}: (V_g\otimes V_h)\otimes V_k\to V_g\otimes (V_h\otimes V_k)$ may fail to satisfy the pentagon equation. The $\cc{O}_4$-obstruction $j^{\omega,\gamma,p}$ exactly measures the failure of the pentagon equation. Namely $j^{\omega,\gamma,p}([g],[h],[k],[l])=d\alpha(g,h,k,l)$. Therefore the $\cc{O}_4$-obstruction is always trivialized in $\cc{H}^4[G,\bC^\times]$. Conversely, take any $\pi\in \cc{H}^4[G/N,\bC^\times]$ such that $p^*(\pi)=d\alpha$. Then $\alpha|_N$ is a 3-cocycle of $N$, and $\alpha$ is ``almost" a lift of $\alpha|_N$ to $G$ except that it is obstructed by $\pi\in \cc{H}^4[G/N,\bC^\times]$. This means $\alpha$ may be constructed from a quadruple $(\omega,\gamma,p,\beta)$(with $\omega=\alpha|_N$) satisfying $d_2\omega=0$, $i^{\omega,\gamma}=0$, and $\pi=j^{\omega,\gamma,p}d\beta$. Therefore the following sequence is exact.
\begin{align}
    G\mathbf{CrExt}(\vc,N)\xrightarrow{\cc{O}_4} \cc{H}^4[G/N,\bC^\times]\xrightarrow{p^*}\cc{H}^4[G,\bC^\times]
\end{align}

A class $\omega\in \cc{H}^3[G,\bC^\times]$ is an extension of $\omega|_N$ to $G$. This means a $\omega\in \cc{H}^3[G,\bC^\times]$ gives a class $\omega|_N\in\cc{H}^3[N,\bC^\times]$ together with a 3-group morphism $G/N\to \underline{\underline{\mathbf{BrPic}}}(\vc_N^{\omega|_N})$. By truncating this 3-group morphism to a 2-group morphism we obtain an element of $  G\mathbf{CrExt}(\vc,N)$. This defines a group homomorphism
\begin{align}
    \cc{R}: \cc{H}^3[G,\bC^\times]\to   G\mathbf{CrExt}(\vc,N).
\end{align}

This can be understood as an restriction map: a class $\alpha\in \cc{H}^3[G,\bC^\times]$ defines a $G$-crossed braided fusion category $\vc_G^\alpha$. Then by restricting to the normal subgroup $N$ we obtain an element of $G\mathbf{CrExt}(\vc,N)$. Another way of understanding it is to view a class $\alpha\in \cc{H}^3[G,\bC^\times]$ as its restriction $\alpha|_N$ plus extension data $\gamma,p,\beta$ that specifies how $\alpha|_N$ is extended to $\alpha$. Then $\cc{R}$ simply forgets the last piece $\beta$, and $\alpha|_N,\gamma,p$ determine an element of $ G\mathbf{CrExt}(\vc,N)$. An element in $ G\mathbf{CrExt}(\vc,N)$ has trivial $\cc{O}_4$-obstruction exactly when it can be lifted to certain $\vc_G^\alpha$. This means the following sequence is exact.
\begin{align}
    \cc{H}^3[G,\bC^\times]\xrightarrow{\cc{R}} G\mathbf{CrExt}(\vc,N)\xrightarrow{\cc{O}_4}\cc{H}^4[G/N,\bC^\times].
\end{align}

Take a class $\alpha\in \cc{H}^3[G,\bC^\times]$, which we view as $\alpha|_N$ plus extension data $\gamma,p,\beta$. Then $\alpha$ lies in the kernel of $\cc{R}$ if and only if $\alpha|_N,\gamma,p$ are trivial. This means $\alpha$ comes from a class $\beta\in \cc{H}^3[G/N,\bC^\times]$ in the LHS spectral sequence. In other words, the following sequence is exact.
\begin{align}
    \cc{H}^3[G/N,\bC^\times]\xrightarrow{p^*}  \cc{H}^3[G,\bC^\times]\xrightarrow{\cc{R}} G\mathbf{CrExt}(\vc,N)
\end{align}

Put together, we have the following.
\begin{theorem}
    There is an exact sequence
    \begin{align}
          \cc{H}^3[G/N,\bC^\times]\xrightarrow{p^*_3}  \cc{H}^3[G,\bC^\times]\xrightarrow{\cc{R}} G\mathbf{CrExt}(\vc,N)\xrightarrow{\cc{O}_4} \cc{H}^4[G/N,\bC^\times]\xrightarrow{p^*_4}\cc{H}^4[G,\bC^\times].\label{eq: exact_seq_1}
    \end{align}
\end{theorem}
This is the critical tool that we will use to compute the set of minimal condensable algebras. As a set we have 
\begin{align}
    G\mathbf{CrExt}(\vc,N)=\ker(\cc{O}_4)\times \imm(\cc{O}_4)=\imm(\cc{R})\times \ker(p_4^*)=\cc{H}^3[G,\bC^\times]/\imm(p^*_3)\times \ker(p_4^*)\label{eq: gSPT_classifiaction}
\end{align}
 
 Therefore the problem reduces to group cohomology calculation.
 \subsection{Physical interpretation}
 Here we provide a simple physical interpretation of the exact sequence Eq.~\eqref{eq: exact_seq_1}. If we accept that minimal condensable algebras are dual to 2+1-D $G$-gSPTs(which we will prove in Sec.~\ref{sec: physics_of_gSPT}), then the exact sequence may be written in physical terms as 
 \begin{align}
     \mathbf{SPT}_{3D}^{G/N}\to\mathbf{SPT}^G_{3D}\to \mathbf{gSPT}^{(G,N)}_{3D}\to \mathbf{SPT}_{4D}^{G/N}\to \mathbf{SPT}_{4D}^{G}.
 \end{align}
 
Here $\mathbf{gSPT}^{(G,N)}_{3D}$ stands for 2+1-D $G$-gSPTs with gapped symmetry $N$ and the gapless sector transforms faithfully under $G/N$. Consider the right half of the sequence. For a $G$-gSPT, the IR symmetry $G/N$ may be anomalous, this anomaly is the image in $\mathbf{SPT}^{G/N}_{4D}$. However this anomaly can not be arbitrary: it must be emergent, meaning that it trivializes when pulled back to $G$. Conversely, for every such emergent anomaly, one can construct a $G$-gSPT using the Wang-Wen-Witten construction~\cite{Wang_2018}. This means the right half of the sequence is exact. 

Take a 2+1-D $G$-SPT, we can tune to a critical point where a quotient $G/N$ becomes critical. This type of gSPT is called weak or non-intrinsic. This defines the map from $\mathbf{SPT}_{3D}^G$ to $\mathbf{gSPT}_{3D}^{G,N}$. Clearly a weak gSPT has no emergent anomaly: we can tune back to the $G$-SPT symmetrically without introducing topological order, which means the gapless sector has non-anomalous $G/N$-action. Conversely if a $G$-gSPT has no emergent anomaly, then its gapless sector can be gapped out without breaking symmetry or introducing topological order. This shows the sequence is exact at the middle term.

Finally we look at the left half of the sequence. If a $G$-SPT comes from a $G/N$-SPT via pull-back, then as we tune to the $G/N$-critical point, the $G/N$-SPT information should all be lost, therefore a weak gSPT coming from a $G/N$-SPT should be equivalent to a trivial gSPT(whatever a trivial gSPT means). The other direction is: if a $G$-SPT becomes a trivial gSPT when tuned to $G/N$-critical point, then it must come from a $G/N$-SPT. This is more subtle as it requires a precise definition of a ``trivial $G$-gSPT". We can only provide an explanation using the decorated domain wall picture, which is discussed in Sec.~\ref{sec: physics_of_gSPT}.

From Eq.~\eqref{eq: gSPT_classifiaction}, we see that the classification of $G$-gSPT has two factors: 
\begin{align}
    \mathbf{gSPT}_{3D}^{G,N}=\underbrace{\cc{H}^3[G,\bC^\times]/\imm(p^*_3)}_{\text{weak gSPT}}\times \underbrace{\ker(p_4^*)}_{\text{emergent anomaly}}
\end{align}

The second factor determines the image of the gSPT in $\cc{H}^4[G/N,\bC^\times]$, i.e. the emergent anomaly of the gSPT. The first factor means that for a fixed emergent anomaly, different $G$-gSPTs are related by stacking with $G$-SPTs. However stacking with a $G$-SPT does not change the gSPT class if the $G$-SPT comes from a $G/N$-SPT. Therefore for a fixed emergent anomaly, the set of $G$-gSPTs is a torsor over $\cc{H}^3[G,\bC^\times]/\imm(p^*_3)$. The first factor can be called the weak gSPT factor-- it is the classification of weak gSPTs.

\section{Physics of 2+1-D $G$-gSPT: decorated domain wall and spectral sequence\label{sec: physics_of_gSPT}}
In the previous section, we defined a class of condensable $E_2$-algebras in $\Z_1[2\vc_G]$, named minimal condensable algebras, and  claimed they are dual to 2+1-D $G$-gSPTs. The mathematical structure of minimal condensable algebras does seem to resemble some physical properties of $G$-gSPTs, as discussed above. However, to rigorously establish a correspondence, we need to develop a pure 2+1-D theory for 2+1-D $G$-gSPTs, which is the goal of this section. In this section we give a pure 2+1-D analysis for 2+1-D $G$-gSPTs, and confirm that they are described by exactly the same structure as minimal condensable algebras. Our tool is the decorated domain wall construction.\footnote{The discussion here is inspired by a simialr analysis in~\cite{ma2025topologicalphasesaveragesymmetries,Ma_2023} for average SPTs(ASPT). In fact ASPT and gSPT are expected to be in 1-1 correspondence~\cite{ma2025topologicalphasesaveragesymmetries}. }

Let us start by reviewing the standard decorated domain wall construction for gapped (bosonic)SPTs. The discussion here follows~\cite{ma2025topologicalphasesaveragesymmetries,wang2021domainwalldecorationsanomalies}. If the protecting symmetry $G$ sits in a group extension 
\begin{align}
    1\to N\to G\to K\to 1
\end{align}
Then we may construct an SPT of $G$ by decorating defects of $K$(domain walls, junctions between domain walls, etc.) with SPTs of $N$. We start with a $K$-broken gapped trivial state, then then we put SPTs of $N$ on defects of $K$, finally we proliferate the $K$-domain walls. This process yields a gapped SPT of $G$ if and only if the conditions for the decoration pattern hold:
\begin{itemize}
    \item The defects of $K$ in each dimension have no t' Hooft anomaly of $N$. This makes sure condensing the domain walls of $K$ results in a gapped state with no ground state degeneracy.
    \item The charge of $N$ is preserved under a continuous deformation of  $K$-defect configuration. This makes sure the resulting state preserves the symmetry $G$.
    \item There is no berry phase accumulated under a closed continuous deformation of defect configuration. 
\end{itemize}

Mathematically, the decorated domain wall construction is described by the Atiyah-Hirzebruch spectral sequence, whose $E_2$-page is 
\begin{align}
    E_2^{p,q}=\cc{H}^p(K,h^q(N)),
\end{align}
where $h^q(N)$ is the classification of invertible phases in $q$ spacetime dimension with symmetry $N$, and $p+q=d+1$ is the spacetime dimension. If $N$ is finite and unitary, then in low dimensions $h^q(N)\simeq \cc{H}^q[N,\bC^\times]$ and the spectral sequence reduces to the LHS spectral sequence.  Physically the term $\cc{H}^p(K,h^q(N))$ describes decorating $N$-invertible states on codimension-$p$ defects of $K$. The condition that the $K$-defects are free of $N$-anomaly leads to the coycle condition. Not all terms on the $E_2$-page describe valid decoration pattern. For a decoration pattern $\omega^{p,q}\in E_2^{p,q}$ to correspond to a valid $G$-SPT, there are certain obstructions that must vanish. These obstructions are described by the differentials
\begin{align}
    d_r: E_r^{p,q}\to E_r^{p+r,q-r+1}.
\end{align}
 For example, for a term $\omega^{0,3}\in \cc{H}^0(K,h^3(N))$ to give rise to a valid 2+1-D $G$-SPT, we must have $d_2(\omega^{0,3})=1$ in $E_2^{2,2}$, $d_3(\omega^{0,3})=1$ in $E_3^{3,1}$ and $d_4(\omega^{0,3})=1$ in $E_4^{4,0}$. 

These obstructions correspond to the conditions for the decoration pattern discussed above:
\begin{itemize}
    \item For $r\le d-p$, $d_r(\omega)=0$ means the defects of $K$ carry no anomaly of $N$.
    \item For $r=d-p$, $d_r(\omega)=0$ means the charge of $N$ is preserved under a continuous deformation of $K$-defects.
    \item For $r=d$, $d_r(\omega)=0$ means there is Berry phase accumulated under closed continuous deformation of $K$-defects.
\end{itemize}

Similar to our discussion in Sec.~\ref{sec: LHS}, when these conditions are met, each condition $d_r\omega=0$ means certain equation has a solution, and we need to choose a solution. This solution will then define a consistent decoration pattern therefore a $G$-SPT.
\hide{Cochain level expression for the differentials $d_i$ in low dimensions can be found in~\cite{}. In dimension 2+1-D, the domain wall decoration pattern and the consistency conditions can be expressed as terms $F^{p,q}\in C^p[K,C^q[N,U(1)]]$, such that 
\begin{align}
    &\delta_0 F^{0,3}=1,\\
    &\delta_1 F^{0,3}\delta_0 F^{1,2}=1\\
    &\delta_2 F^{0,3}\delta_1 F^{1,2}\delta_0 F^{2,1}=1\\
    &\delta_3 F^{0,3}\delta_2 F^{1,2}\delta_1 F^{2,1}\delta_0 F^{3,0}=1\\
    &\delta_4 F^{0,3}\delta_3 F^{1,2}\delta_2 F^{2,1}\delta_1 F^{3,0}=1.
\end{align}
The explicit form of the cochain-level differentials $\delta_i: C^p[K,C^q[N,U(1)]]\to C^{p-i}[K,C^{p+i+1}[N,U(1)]]$ can be found in Appendix~\ref{}.
}

Now we consider a 2+1-D gapless SPT with exact symmetry $G$ and gapped symmetry $N$. Although the system is gapless, the symmetry $N$ only acts on gapped degrees of freedom, and it is still legitimate to talk about the SPT/SET class of $N$. This class is a topological invariant of the system, in the sense that it will not change as long as the $N$-charge gap is not closed. If the $N$-charges form a nontrivial topological order, then it should be considered as a gapless SET. Furthermore, we can talk about the gapped $N$-phases on defects of $G$. These are also topological invariants of the system. These $N$-phases must be gapped and short-ranged entangled(otherwise we call it a gSET). Therefore for any gSPT we can identify a domain wall decoration pattern. This domain wall decoration pattern is exactly the topological skeleton of the gSPT. This decoration pattern must still satisfy the first two conditions for gapped SPTs: 
\begin{itemize}
    \item Since $N$ is gapped and not broken, there should be no $N$-anomaly on defects of $K$.
    \item Since $N$ is not broken, the charge of $N$ should be preserved under a continuous deformation of $K$-defect configuration.
\end{itemize}

However, we no longer require the last condition to hold. This is because the $K$-action on the gapless state can be anomalous. This means upon a closed deformation of $K$-defects, the $N$-SPTs decorated on $K$-defects may contribute a Berry phase $\cc{O}$, and the anomaly of the $K$-action on the gapless state contributes a phase $\Omega$. Then the wavefunction of the gapless state is single valued as long as $\cc{O}=\Omega^{-1}$. This is exactly the anomaly cancellation mechanism for igSPTs: the anomaly from the gapped sector cancels the anomaly of the gapless sector. We summarize the above discussion as the following physical theorem. 

\begin{ptheorem}
    A 2+1-D $G$-gSPT is described by a term $\omega\in\cc{H}^3[N,\bC^\times]$ such that $d_2\omega=0, d_3\omega=0$ together with a solution to these equations $(\gamma,p)$. The emergent anomaly of the gSPT is $j^{\omega,\gamma,p}$, where $j^{\omega,\gamma,p}$ is the $\cc{O}_4$-obstruction of lifting $\omega$ to $G$. 
\end{ptheorem}

At this point, it should become clear that 2+1-D $G$-SPTs are exactly dual to minimal condensable algebras in the \symto $\Z_1[2\vc_G]$. Furthermore, our discussion on minimal condensable algebras immediately leads to the following.
\begin{corollary}
    A 2+1-D $G$-gSPT with gapped symmetry $N$ is equivalently described by a $G$-crossed braided fusion structure on $\vc_N$.
\end{corollary}

\begin{corollary}
    The set of 2+1-D $G$-gSPTs with gapped symmetry $N$, denoted as $\mathbf{gSPT}^{(G,N)}_{3D}$, is given by 
    \begin{align}  \mathbf{gSPT}_{3D}^{G,N}=\underbrace{\cc{H}^3[G,\bC^\times]/\imm(p^*_3)}_{\text{weak gSPT}}\times \underbrace{\ker(p_4^*)}_{\text{emergent anomaly}}
    \end{align}
    where $p_3^*,p_4^*$ are the maps on cohomology induced by $p: G\to G/N$.
\end{corollary}
\section{Examples of minimal condensable algebras and $G$-gSPTs\label{sec: gSPT_eg}}
In this section, we apply our general results in Sec.~\ref{sec: minimal_cd_alg}, Sec.~\ref{sec: physics_of_gSPT} to study concrete examples. We compute the classification of 2+1-D gSPTs for several $G$ and $N$, and discuss their emergent anomalies. Our main tool is the formula
\begin{align}
    \mathbf{gSPT}_{3D}^{G,N}=\underbrace{\cc{H}^3[G,\bC^\times]/\imm(p^*_3)}_{\text{weak gSPT}}\times \underbrace{\ker(p_4^*)}_{\text{emergent anomaly}},
\end{align}
where 
\begin{align}
    p_3^*:\cc{H}^3[G/N,\bC^\times]\to \cc{H}^3[G,\bC^\times],
\end{align}
and 
\begin{align}
    p_4^*:\cc{H}^4[G/N,\bC^\times]\to \cc{H}^4[G,\bC^\times],
\end{align}
are induced by $p: G\to G/N$.

When $N=G$, $p_3^*,p_4^*$ are zero maps and we have 
\begin{align}
    \mathbf{gSPT}_{3D}^{G,G}=\cc{H}^3[G,\bC^\times]=\mathbf{SPT}_{3D}^{G}.
\end{align}

When $N=\{e\}$, $p_3^*,p_4^*$ are identity maps and we have
\begin{align}
    \mathbf{gSPT}_{3D}^{G,\{e\}}=0.
\end{align}

Therefore in the following we only consider $N$ to be a proper normal subgroup of $G$.
\subsection{$G=\bZ_4,N=\bZ_2$}

 We need to understand the map on cohomology $p^*_3: \cc{H}^3[\bZ_2,U(1)]\to \cc{H}^3[\bZ_4,U(1)]$. Notice 
    \begin{align*}
        \cc{H}^3[\bZ_2,U(1)]=\cc{H}^4[\bZ_2,\bZ]=\<x\cup x\>
    \end{align*}
    where $x$ generates $\cc{H}^2[\bZ_2,\bZ]=\bZ_2$. Similarly $\cc{H}^3[\bZ_4,U(1)]=\<a\cup a\>$, with $a$ generating $\cc{H}^2[\bZ_4,\bZ]=\bZ_4$. The map $p^*$ maps $x$ to $2a$, therefore  $x\cup x$ to $4a\cup a=0$. We see that $p_3^*=0$. Since $H^4[\bZ_n,U(1)]=0$, we also have $p_4^*=0$. Therefore we have 
    \begin{align}  \mathbf{gSPT}_{3D}^{\bZ_4,\bZ_2}=\underbrace{\bZ_4}_{\text{weak gSPT}}\times \underbrace{0}_{\text{emergent anomaly}}
    \end{align}
    $\cc{H}^3[(\bZ_4,\bZ_2),U(1)]\simeq \cc{H}^4[\bZ_4,U(1)]=\bZ_4$.   Physically, this means 
    \begin{enumerate}
        \item There is no igSPT for $G=\bZ_4,N=\bZ_2$ since $\cc{H}^4[G/N,U(1)]=0$.
        \item All four SPTs of $\bZ_4$ become four different weak gSPTs when $G/N=\bZ_2$ is made critical.
    \end{enumerate}

\subsection{$G=\bZ_n,~N=\bZ_p$}
Let $n=pq$, and $N=\bZ_p$. Since $\cc{H}^4[\bZ_n,U(1)]=0$, there can not be an emergent anomaly and we only need to understand the map $p_3^*: \cc{H}^3[\bZ_q,U(1)]\to \cc{H}^3[\bZ_n,U(1)]$.

We notice that $\cc{H}^3[\bZ_n,U(1)]\simeq \cc{H}^4[\bZ_n,\bZ]$ is generated by $a\cup a$, with $a$ generating $\cc{H}^2[\bZ_n,\bZ]$. Similarly $\cc{H}^3[\bZ_q,U(1)]\simeq \cc{H}^4[\bZ_q,\bZ]$ is generated by $x\cup x$ with $x$ generating $\cc{H}^2[\bZ_q,\bZ]$. $p^*$ maps $x$ to $pa$, therefore maps $x\cup x$ to $p^2 a\cup a$. We have 
\begin{align*}
    p_3^*:  \cc{H}^3[\bZ_q,U(1)]\simeq \bZ_q\xrightarrow{\times p^2} \bZ_n\simeq \cc{H}^3[\bZ_n,U(1)]
\end{align*}

Therefore $\imm p_3^*=\bZ_{\frac{n}{(n,p^2)}}=\bZ_{\frac{q}{(p,q)}}$ and $\cc{H}^3[\bZ_n,U(1)]/\imm p_3^*\simeq \bZ_{p(p,q)}$. We see that
 \begin{align}  \mathbf{gSPT}_{3D}^{\bZ_n,\bZ_p}=\underbrace{\bZ_{p(p,q)}}_{\text{weak gSPT}}\times \underbrace{0}_{\text{emergent anomaly}}
    \end{align}

In total, the number of 2+1-D $\bZ_n$-gSPTs is
    \begin{align}
        \sum_{p|n,p\neq n}p(p,q)
    \end{align}
 
All of them are weak gSPTs. However, the restriction map $\cc{R}$ is not an isomorphism in general: different SPTs can become identical gSPT when tuned to critical points.
\subsection{$G=\bZ_2\times \bZ_4,~ N=\bZ_2=\<0,2\>$}
The general case $\bZ_{n_1}\times \bZ_{n_2}$ will be treated in the next subsection. As a teaser, here we consider the case where $N$ is the $\bZ_2$ subgroup of the $\bZ_4$ factor, and $G/N=\bZ_2\times \bZ_2$. 

The relevant cohomology groups are $\cc{H}^4[\bZ_2\times \bZ_4,U(1)]\simeq \cc{H}^4[\bZ_2\times \bZ_2,U(1)]=\bZ_2^2$, $\cc{H}^3[\bZ_2\times \bZ_4,U(1)]=\bZ_2\times \bZ_4\times \bZ_2$ and $\cc{H}^3[\bZ_2\times \bZ_2,U(1)]=\bZ_2^3$. The cohomology ring of $\bZ_2\times \bZ_2$ and $\bZ_2\times \bZ_4$ can be found in~\cite{cohomology_ring}. We have 
\begin{align*}
    \cc{H}^{*}[\bZ_2\times \bZ_4,\bZ]\simeq \bZ[a,b,c]/\<2a,4b,2c\>
\end{align*}
with $|a|=|b|=2, |c|=3$. In degree $\le 6$, we also have 
\begin{align*}
    \cc{H}^{*\le 6}[\bZ_2^2,\bZ]\simeq \bZ[x,y,z]/\<2x,2y,2z\>.
\end{align*}
with $|x|=|y|=2, |z|=3$. The projection $p: \bZ_2\times \bZ_4\to \bZ_2\times \bZ_2$ induces a map on cohomology that maps $x\mapsto a~y\mapsto 2b,~ z\mapsto c$. Therefore we have 
\begin{align*}
    &\imm p_3^*=\<a^2\>\simeq \bZ_2,~\cc{H}^3[\bZ_2\times \bZ_4,U(1)]/\imm p_3^*\simeq \bZ_2\times \bZ_4,\\
    &\ker p_4^*=\<y\cup z\>\simeq \bZ_2,
\end{align*}
from which we deduce

 \begin{align}  \mathbf{gSPT}_{3D}^{\bZ_2\times \bZ_4,\<(0,2)\>}=\underbrace{\bZ_{2}\times\bZ_4}_{\text{weak gSPT}}\times \underbrace{\bZ_2}_{\text{emergent anomaly}}
    \end{align}

The second factor $\ker p_4^*=\bZ_2$ means that there are igSPTs: the element $y\cup z\in \cc{H}^4[\bZ_2\times \bZ_2,U(1)]$ is trivialized when pulled back to $\bZ_2\times \bZ_4$, thus can be a valid emergent anomaly. Fixing an emergent anomaly(either trivial or $y\cup z$), different gSPTs are related by stacking with $\bZ_2\times \bZ_4$-SPTs. However not all different $\bZ_2\times \bZ_4$-SPTs give different gSPTs by stacking. Specifically, stacking with the SPT protected purely by the $\bZ_2$ factor does not affect the gSPT class.

In conclusion, there are  $16$ gSPTs with symmetry $\bZ_2\times \bZ_4$ and gapped symmetry $N=\bZ_2=\<(0,2)\>$, 8 of which are igSPTs.
\subsection{ $G=\bZ_{n_1}\times \bZ_{n_2},~ N=\bZ_p=\<(0,q)\>$}
We may assume  $n_1|n_2$ are the elementary divisors of the abelian group. Let $n_2=pq$ and $N=\bZ_p=\<(0,q)\>$, $G/N=\bZ_{n_1}\times \bZ_q$. The integer cohomology rings $\cc{H}^*$ in degree $\le 6$ are
\begin{align*}
    &\cc{H}^{*\le 6}[\bZ_{n_1}\times \bZ_{n_2},\bZ]\simeq \bZ[a,b,c]^{\le 6}/\<n_1a,n_2b,n_1c\>\\
    &\cc{H}^{*\le 6}[\bZ_{n_1}\times \bZ_q,\bZ]\simeq \bZ[x,y,z]^{\le 6}/\<n_1x,qy,(n_1,q)z\>.
\end{align*}
with $|a|=|b|=|x|=|y|=2,~|c|=|z|=3$.  $p^*$ maps $x\mapsto a, y\mapsto pb, z\mapsto \frac{n_1}{(n_1,q)}c$. Thus on $\cc{H}^4$ we have $x^2\mapsto a^2, y^2\mapsto p^2 b^2, x\cup y\mapsto p a\cup b$, on $\cc{H}^5$ we have $x\cup z\mapsto \frac{n_1}{(n_1,q)}a\cup c, y\cup z\mapsto \frac{pn_1}{(n_1,q)}b\cup c$. 

From this one can deduce that 
\begin{align*}
    \cc{H}^3[\bZ_{n_1}\times \bZ_{n_2},U(1)]/\imm p_3^*\simeq \bZ_{p(p,q)}\times \bZ_{(n_1,p)},
\end{align*}
and 
\begin{align*}
    \ker p_4^*=\bZ_{(p,q,n_1))}.
\end{align*}

Therefore gSPTs with symmetry $\bZ_{n_1}\times \bZ_{n_2}$(with $n_1|n_2$) and $N=\bZ_p=\<(0,q)\>$ have classification 
 \begin{align}  \mathbf{gSPT}_{3D}^{\bZ_{n_1}\times \bZ_{n_2},\bZ_p=\<(0,q)\>}=\underbrace{\bZ_{p(p,q)}\times \bZ_{(n_1,p)}}_{\text{weak gSPT}}\times \underbrace{\bZ_{(p,q,n_1)}}_{\text{emergent anomaly}}.
    \end{align}

We see that there exists 2+1-D igSPT with symmetry $\bZ_{n_1}\times \bZ_{n_2}$(with $n_1|n_2$) and $N=\bZ_p=\<(0,q)\>$ if and only if $(p,q,n_1)\neq 1$.
\subsection{$G=S_3,~N=\bZ_3$}
Let $S_3=\<r,s|r^3=s^2=1,srs=r^2\>$. The only proper normal subgroup is $N=\<r\>=\bZ_3$, and we have $G/N=\bZ_2$. Since $\cc{H}^4[G/N,U(1)]=0$, there is no emergent anomaly thus no igSPT. We next classify weak gSPTs. We need to understand the map $p_3^*: \cc{H}^3[\bZ_2,U(1)]\to \cc{H}^3[S_3,U(1)]$. Consider the LHS spectral sequence $\cc{H}^p[\bZ_2,\bZ^q[\bZ_3,U(1)]]\To \cc{H}^{p+q}[S_3,U(1)]$. For $p+q=3$ we only have two nonzero terms on the $E_2$-page: $\cc{H}^3[\bZ_2,U(1)]\simeq \bZ_2$ and $\cc{H}^0[\bZ_2,\cc{H}^3[\bZ_3,U(1)]]=\bZ_3$. We know that $\cc{H}^3[S_3,U(1)]=\bZ_6$. Therefore all terms in $\cc{H}^3[\bZ_2,U(1)]$ survive to $\cc{H}^3[S_3,U(1)]$. Thus $\imm p_3^*=\bZ_2$ and $\cc{H}^3[S_3,U(1)]/\imm p_3^*=\bZ_3$. We conclude

 \begin{align}  \mathbf{gSPT}_{3D}^{S_3,\bZ_3}=\underbrace{\bZ_3}_{\text{weak gSPT}}\times \underbrace{0}_{\text{emergent anomaly}}
    \end{align}
In conclusion, there are 3 gSPTs with $G=S_3, N=\bZ_3$, and all of them are weak gSPTs. 
\subsection{$G=D_4, N=Z(D_4)=\bZ_2$}
Let $D_4=\<r,s|r^4=s^2=(rs)^2=1\>$. There are in total 4 nontrivial normal subgroups. Let us consider the case that can potentially have an emergent anomaly: $N=\bZ_2=\<r^2\>$, and $G/N=\bZ_2^2$. We are interested in maps $p_3^*: \cc{H}^3[\bZ_2^2,U(1)]\simeq \bZ_2^3\to \cc{H}^3[D_4,U(1)]\simeq \bZ_2^2\times \bZ_4$, and $p_4^*: \cc{H}^4[\bZ_2^2,U(1)]\simeq \bZ_2^2\to \cc{H}^4[D_4,U(1)]\simeq \bZ_2^2$. We again use the LHS spectral sequence. Let $\cc{H}^*(\bZ_2^2,\bZ)=\bZ[a,b,c]/\<\text{relations}\>$ with $|a|=|b|=2,|c|=3$. It turns out $a\cup b\in \cc{H}^4[\bZ_2^2,\bZ]\simeq \cc{H}^3[\bZ_2^2,U(1)]$ is hit by a differential($d_2$) from $\cc{H}^1[\bZ_2^2,\cc{H}^1[\bZ_2,U(1)]]$, and $\imm p_3^*=\bZ_2^2$. Therefore $\cc{H}^3[D_4,U(1)]/\imm p_3^*\simeq \bZ_4$. For $p_4^*$, it turns out all elements in $\cc{H}^4[\bZ_2^2,U(1)]$ are hit by a differential from $\cc{H}^2[\bZ_2^2,\cc{H}^1[\bZ_2,U(1)]]$, therefore $\ker p_4^*=\bZ_2^2$.  We conclude
\begin{align}
\mathbf{gSPT}_{3D}^{D_4,Z(D_4)}=\underbrace{\bZ_4}_{\text{weak gSPT}}\times \underbrace{\bZ_2^2}_{\text{emergent anomaly}}
\end{align}

There are three different types of nontrivial emergent anomalies, all of which are trivialized in $D_4$ by a $d_2$ differential.

\section{2+1-D Gapless SPT with generalized symmetry\label{sec: gSPT_2}}
In this section we apply the \symto club-sandwich construction to construct 2+1-D gapless SPTs with generalizes symmetries. For a symmetry category $\e{C}$, this amounts to choosing to non-Lagrangian condensable $E_2$-algebra that preserves the symmetry $\e{C}$. In this section we will only consider ``simple" condensable algebras, meaning that the higher structures of the algebra such as the associator, the braiding, are taken to be trivial, and we will only define condensable algebras at the object level.
\subsection{igSPT with 2-group symmetry $2\vc_{(\bZ_2^{[1]}\times \bZ_2^{[1]})\rtimes \bZ_2^{[0]}}$}
We consider a \symto $\Z_1[2\vc_{D_4}]$. We take $D_4=\<r,s|r^4=s^2=(rs)^2=1\>$. Some properties of this topological order are summarized below.
\begin{itemize}
    \item The gauge charges form a symmetric fusion category $\rep(D_4)$. There are six simple objects.
    \begin{enumerate}
        \item Four 1-dimensional representations, with $r=\pm, s=\pm$. The corresponding gauge charges are denoted as $B_{\pm,\pm}$. 
        \item One 2-dimensional representation, with 
        \begin{align}
            s=\begin{pmatrix}
             0 & 1\\
               1 & 0
            \end{pmatrix}, ~r=\begin{pmatrix}
             w & 0\\
               0 & w^3
            \end{pmatrix}
        \end{align}
        where $w=e^{2\pi i/4}$. The corresponding non-abelian gauge charge is denoted as $Q$. They have fusion rule
        \begin{align}
            B_{\pm,\pm}\otimes Q=Q\otimes B_{\pm,\pm}=Q,~ Q\otimes Q=B_{+,+}\oplus B_{+,-}\oplus B_{-,+}\oplus B_{-,-}.
        \end{align}
    \end{enumerate}
    \item There are five flux sectors labeled by the five conjugacy classes: $1, M_{\{s,sr^2\}}, M_{r^2},M_{\{r,r^3\}},M_{\{sr,sr^3\}}$.
\end{itemize}

On the symmetry boundary, we choose gapped minimal boundary $\fk{C}[D_4,\bZ_2^2,0,0]$, where $\bZ_2^2=\<s,r^2\>$. By Example.~\ref{eg: fusion 2-cat_2}, the symmetry category is the 2-group symmetry $2\vc_{(\bZ_2^{[1]}\times \bZ_2^{[1]})\rtimes \bZ_2^{[0]}}$, where the 0-form symmetry exchanges the two $\bZ_2$-factors of the 1-form symmetry. We can derive this symmetry category by analyzing the corresponding Lagrangian condensation.

 The Lagrangian condensation on the symmetry boundary can be decomposed into two steps. In the first step the condensable $E_2$-algebra is $\mathbf{Func}[D_4/(\bZ_2\times \bZ_2),\vc]=\Sigma\bC[D_4/(\bZ_2\times \bZ_2)]=\Sigma(1\oplus B_{-+})$. This means the charge $B_{-+}$ is condensed. This condensation reduces the $D_4$-gauge theory to a $\bZ_2^2$-gauge theory. In particular, we have:
 \begin{itemize}
     \item  The charge $B_{+-}$ stays nontrivial and becomes a gauge charge of the $\bZ_2^2$-gauge theory. The charge $Q$ splits into two irreducible representations of $\bZ_2\times \bZ_2=\<s,r^2\>$: $Q\to q_{1}\oplus q_2$. Together with the trivial charge they form the fusion category $\rep(\bZ_2\times \bZ_2)$. 
     \item The fluxes $M_{\{r,r^3\}},~M_{\{sr,sr^3\}}$ braid nontrivially with $B_{+-}$ and get confined by the condensation. The flux $M_{r^2}$ and $M_{\{s,sr^2\}}$ stay deconfined. Furthermore $M_{\{s,sr^2\}}$ splits into two fluxes of $\bZ_2^2$: $M_{\{s,sr^2\}}\to M_s\oplus M_{sr^2}$. 
 \end{itemize}
 
 In the second step, we condense the fluxes $M_{r^2}, M_s$. This is described by the Lagrangian algebra $\vc_{\bZ_2\times \bZ_2}$ in $\Z_1[2\vc_{\bZ_2\times \bZ_2}]$. In total, the charge $B_{-+}$ and the fluxes $M_{\{s,sr^2\}}, M_{r^2}$ are condensed.
 
 \hide{The topological operators on the symmetry boundary are as follows.
 \begin{itemize}
     \item Since $M_{\{s,sr^2\}}, M_{r^2}$ are condensed, the only nontrivial flux on the symmetry boundary is $M_{\{r,r^3\}}$. Furthermore, it decomposes into two copies of a simple object $M_{r,r^3}\to M'_r\oplus M'_r$ on the symmetry boundary. Therefore the symmetry fusion 2-category has two connected components.
     \item Since the charge $B_{-+}$ is the condensed, the charges $B_{+-}, Q$ are nontrivial on the symmetry boundary. Furthermore the charge $Q$ splits into two charges $Q_1, Q_2$. Together they generate a 1-form symmetry $\bZ_2^{[1]}\times \bZ_2^{[1]}$. Including the condensation descendants, they generate a fusion 2-category $2\rep(\bZ_2\times \bZ_2)$, which is the component of the full symmetry category connected to the unit.
     \item The two $\bZ_2^2$-charges $Q_1,Q_2$ are exchanged under the action of the $\bZ_2$ 0-form symmetry.
 \end{itemize}
This analysis shows the symmetry category defined by the Lagrangian algebra $\e{A}[\bZ_2^2,1]$ is indeed a 2-group symmetry $2\vc_{(\bZ_2^{[1]}\times \bZ_2^{[1]})\rtimes \bZ_2^{[0]}}$.
}
We next construct an igSPT with this 2-group symmetry by choosing a non-Lagrangian condensable algebra in $\Z_1[2\vc_{D_4}]$. Consider the condensable $E_2$-algebra $\mathbf{Func}[D_4/\bZ_4,\vc]$. This is a pure charge condensation that Higgses the gauge field to $\bZ_4=\<r\>$. We have $\mathbf{Func}[D_4/\bZ_4,\vc]=\Sigma\bC[D_4/\bZ_4]=\Sigma(1\oplus B_{+-})$. I.e. the charge $B_{+-}$ is condensed. This condensation reduces the $D_4$-gauge theory to a $\bZ_4$-gauge theory. In particular, we have 
\begin{itemize}
    \item The charges $B_{-+}$ and $Q$ stay nontrivial. Since $r=-1$ in $B_{-+}$, $B_{-+}$ becomes the double charge $e^2$ of the $\bZ_4$-gauge theory. The non-abelian charge $Q$ splits into two irreducible representations of $\bZ_4$: $Q\to e+e^3$. 
    \item The flux $M_{\{s,sr^2\}}$ and $M_{\{sr,sr^3\}}$ braid nontrivially with $B_{+-}$ and get confined. The fluxes $M_{r^2}, M_{r,r^3}$ stay deconfined. Furthermore $M_{r,r^3}$ splits: $M_{r,r^3}\to m_r+m_{r^3}$. These are the fluxes of the $\bZ_4$-gauge theory.
\end{itemize}
\begin{figure}[h]
    \centering
        \begin{tikzpicture}[baseline=(current  bounding  box.center)]
  \node[label,below] at (3,0){$\fk{M}$};
  \node[label,below] at (0,0){$\fk{B}_\phys$};
\fill[zima!70] (0,0) rectangle (3,4);
\fill[zima] (3,0) rectangle (6,4);
        \draw[thick] (3,0) -- (3,4);
          \draw[thick] (6,0) -- (6,4);
          \draw[thick] (0,0) -- (0,4);
\node[label,above] at (3,4){$\<B_{+-}\>$};
\node[label,above] at (6,4){$\<B_{-+}, M_{\{s,sr^2\}},M_{r^2}\>$};
\node[label] at (1.5,2){$\Z_1[2\vc_{\bZ_4}]$};
\node[label] at (4.5,2){$\Z_1[2\vc_{D_4}]$};
\node[label] at (2.6,1){$e+e^3\leftarrow Q$};
\node[label] at (3.2,0.4){$e^2\leftarrow B_{-+}$};
\node[label] at (2.8,2.9){$m+m^3\leftarrow M_{r,r^3}$
};
\node[label] at (3.1,3.5){$m^2\leftarrow M_{r^2}$};
\node[label,below] at (6,0){$\fk{B}_\sym$};
\node[label] at (7,2){$\simeq$};
\fill[zima!70] (8,0) rectangle (11,4);
\draw[thick] (8,0) -- (8,4);
\draw[thick] (11,0) -- (11,4);
\node[label] at (9.5,2){$\Z_1[2\vc_{\bZ_4}]$};
\node[label, above] at (11,4){$\<e^2,m^2\>$};
\node[label, below] at (11,0){$\fk{B}_\sym^\IR$};
\node[label,below] at (8,0){$\fk{B}_\phys$};
\end{tikzpicture}
\caption{The club-sandwich construction of a $(\bZ^{[1]}_2\times\bZ_2^{[1]})\rtimes \bZ_2^{[0]}$-igSPT.}
    \label{fig: 2-group igSPT}
\end{figure}
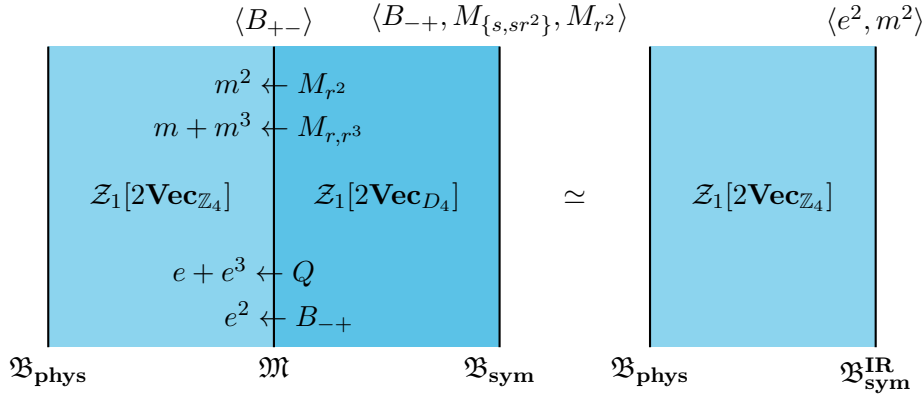

The structure of the club-sandwich is shown in Fig.~\ref{fig: 2-group igSPT}.
We note that this choice of condensable algebra preserves the 2-group symmetry $2\vc_{(\bZ_2^{[1]}\times \bZ_2^{[1]})\rtimes \bZ_2^{[0]}}$: The domain wall $\fk{M}$ and the symmetry boundary $\fk{B}_\sym$ share no common charge or flux condensation. Therefore there is no 0-form nor 1-form symmetry breaking, and the 2-group symmetry is preserved.

Next we analyze the IR symmetry of this $2\vc_{(\bZ_2^{[1]}\times \bZ_2^{[1]})\rtimes \bZ_2^{[0]}}$-gSPT. As explained in Sec.~\ref{sec: gapless}, this is obtained by fusing the domain wall $\fk{M}$ with $\fk{B}_\sym$ to obtain a new symmetry boundary of the condensed phase $\Z_1[2\vc_{\bZ_4}]$. To determine the nature of this new symmetry boundary, we only need to determine the charge and flux condensation on it. See Fig.~\ref{fig: 2-group igSPT}. The charge $e,e^3$ become the non-abelian charge $Q$, while the charge $e^2$ becomes the abelian charge $B_{-+}$. Since on $\fk{B}_\sym$ only the charge $B_{-+}$ is condensed, we see that the effective symmetry boundary of the $\bZ_4$-gauge theory condenses the charge $e^2$. The flux $m,m^3$ of the $\bZ_4$-gauge theory becomes the flux $M_{\{r,r^3\}}$ when moved into the $D_4$-theory, and the flux $m^2$ becomes the flux $M_{r^2}$. Since  on $\fk{B}_\sym$ the flux $M_{r^2}$ is condensed, we see that the effective symmetry boundary of the $\bZ_4$-gauge theory condenses the flux $m^2$.  Therefore the effective gapped boundary for the $\bZ_4$-gauge theory is defined by the property that the charge $e^2$ and the flux $m_{r^2}$ condense. This means the IR symmetry of this gSPT is $\fk{C}[\bZ_4,\bZ_2,0,0]$. According to Example.~\ref{eg: anomalous 2-group}, this symmetry is the  \textit{anomalous} 2-group symmetry 
\begin{align}
    2\vc_{\bZ_2^{[0]}\times \bZ_2^{[1]}}^\omega.
\end{align}

Therefore the above construction yields an igSPT with full symmetry $2\vc_{(\bZ_2^{[1]}\times \bZ_2^{[1]})\rtimes \bZ_2^{[0]}}$ and anomalous IR symmetry $ 2\vc_{\bZ_2^{[0]}\times \bZ_2^{[1]}}^\omega$.
\subsection{igSPT with symmetry $2\vc_{\bZ_4}\times 2\rep(D_4)$}
The \symto is $\Z_1[2\vc_{\bZ_4\times D_4}]=\Z_1[2\vc_{\bZ_4}]\boxtimes \Z_1[2\vc_{D_4}]$. The symmetry boundary is defined by the Lagrangian algebra $\mathbf{Func}[\bZ_4,\vc]\boxtimes \vc_{D_4}$. Physically, all charges of the $\bZ_4$-gauge field are condensed, and all fluxes of the $D_4$-gauge field are condensed. To construct a gSPT with this symmetry, we construct a non-Lagrangian condensable algebra in two steps.  In the first step, we condense all abelian charges of $D_4$: $B_{\pm,\pm}$. This will reduce the gauge group to $\bZ_4\times \bZ_2$, with the second $\bZ_2$-factor generated by the central element $r^2\in D_4$. Specifically,
\begin{itemize}
    \item The non-abelian charge $Q$ remains deconfined and splits into $e_{\bZ_2}\oplus e_{\bZ_2}$, with $e_{\bZ_2}$ the $\bZ_2$-charge. 
    \item The flux $M_{r^2}$ is the only deconfiend flux in the $D_4$-sector and becomes the $\bZ_2$-flux $M_{r^2}\to m_{\bZ_2}$. 
\end{itemize}

In the second step, we condense the flux $m_{\bZ_4}^2m_{\bZ_2}$, where $m_{\bZ_4}$ is the elementary flux of the $\bZ_4$-gauge theory. This reduces the gauge group to $\bZ_4$. Specifically, 
\begin{itemize}
    \item The charge $e_{\bZ_4}e_{\bZ_2}$ stays deconfined, where $e_{\bZ_4}$ is the elementary charge in the $\bZ_4$-gauge theory. It generates the charges of $\Z_1[2\vc_{\bZ_4}]$.
    \item The flux $m_{\bZ_4}$ stays deconfined and generates the fluxes of $\Z_1[2\vc_{\bZ_4}]$. 
\end{itemize}

Combine the above two steps, we obtain a non-Lagrangian condensable algebra in $\Z_1[2\vc_{\bZ_4}]\boxtimes \Z_1[2\vc_{D_4}]$. We notice that this defines a 2+1-D gapless phase that preserves the symmetry $2\vc_{\bZ_4}\boxtimes 2\rep(D_4)$: The domain wall $\fk{M}$ condenses $\{B_{\pm,\pm},M_{r^2}m_{\bZ_4}^2\}$, which has trivial intersection with the condensation on the symmetry boundary. 

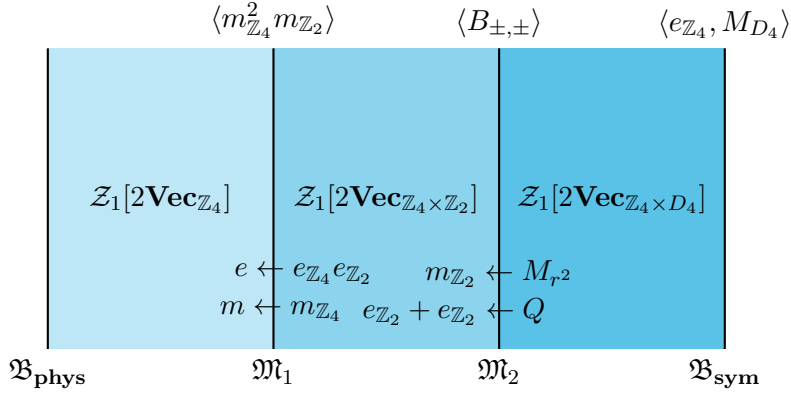
\begin{figure}[h]
    \centering
        \begin{tikzpicture}[baseline=(current  bounding  box.center)]
  \node[label,below] at (3,0){$\fk{M}_1$};
    \node[label,below] at (6,0){$\fk{M}_2$};
  \node[label,below] at (0,0){$\fk{B}_\phys$};
  \node[label,below] at (9,0){$\fk{B}_\sym$};
\fill[zima!40] (0,0) rectangle (3,4);
\fill[zima!70] (3,0) rectangle (6,4);
\fill[zima] (6,0) rectangle (9,4);
        \draw[thick] (3,0) -- (3,4);
          \draw[thick] (6,0) -- (6,4);
          \draw[thick] (0,0) -- (0,4);
          \draw[thick] (9,0) -- (9,4);
\node[label,above] at (3,4){$\< m_{\bZ_4}^2m_{\bZ_2}\>$};
\node[label,above] at (6,4){$\<B_{\pm,\pm}\>$};
\node[label,above] at (9,4){$\<e_{\bZ_4},M_{D_4}\>$};
\node[label] at (1.5,2){$\Z_1[2\vc_{\bZ_4}]$};
\node[label] at (4.5,2){$\Z_1[2\vc_{\bZ_4\times \bZ_2}]$};
\node[label] at (7.5,2){$\Z_1[2\vc_{\bZ_4\times D_4}]$};
\node[label] at (3.4,1){$e\leftarrow e_{\bZ_4}e_{\bZ_2}$};
\node[label] at (3.1,0.5){$m\leftarrow m_{\bZ_4}$};
\node[label] at (5.4,0.5){$e_{\bZ_2}+e_{\bZ_2}\leftarrow Q$};
\node[label] at (6,1){$m_{\bZ_2}\leftarrow M_{r^2}$};
\end{tikzpicture}
\caption{The club-sandwich construction of a $2\vc_{\bZ_4}\boxtimes 2\rep(D_4)$-igSPT. The domain wall $\fk{M}$ is obtained via two steps of condensation.}
    \label{fig: non-invertible igSPT}
\end{figure}

To deduce the IR symmetry of this gSPT, we again fuse the domain wall $\fk{M}$ with $\fk{B}_\sym$ to obtain a new gapped boundary of $\Z_1[2\vc_{\bZ_4}]$. The nature of this gapped boundary is determined by the condensation on it. Observe that $e^2=(e_{\bZ_4}e_{\bZ_2})^2=e_{\bZ_4}^2$ can be moved into the $\Z_1[2\vc_{\bZ_4}]\boxtimes \Z_1[2\vc_{D_4}]$ region and is condensed on $\fk{B}_\sym$.  The flux $m^2\simeq m_{\bZ_4}^2\simeq M_{r^2}$ can be moved into the $\Z_1[2\vc_{\bZ_4}]\boxtimes \Z_1[2\vc_{D_4}]$ region and is also condensed on $\fk{B}_\sym$. Therefore the effective gapped boundary of the $\bZ_4$-gauge theory condenses $e_{\bZ_4}^2$ and $m_{\bZ_4}^2$.  We see that the IR symmetry is again $\fk{C}[\bZ_4,\bZ_2,0,0]$, an anomalous 2-group symmetry. In conclusion, the above construction yields an igSPT with exact non-invertible symmetry $2\vc_{\bZ_4}\boxtimes 2\rep(D_4)$ and an emergent anomalous 2-group symmetry $2\vc_{\bZ_2^{[0]}\times \bZ_2^{[1]}}^\omega$ in the IR.
\vspace{0.5cm}

\textit{Note--} As we were finalizing this manuscript, we learned of the independent work of~\cite{bhardwaj2025gappedphases21dnoninvertible}, which explores gapped 2+1D phases with generalized symmetries and has some overlap with our results.

\textbf{Acknowledgments}

We thank Weicheng Ye, Liang Kong, Zhi-Hao Zhang, Hao Zheng, Sakura Schafer-Nameki for discussions. We thank Lakshya Bhardwaj,  Yuhan Gai, Sheng-Jie Huang,  Kansei Inamura, Sakura Schafer-Nameki, Apoorv Tiwari, Alison WarmanThis for coordinating submission with their upcoming work~\cite{sakura_gapless_21d}. This work was supported by NSERC and the European Commission under the Grant Foundations of Quantum Computational Advantage.
\newpage
\bibliography{main}
\newpage
\appendix
\section{Some useful higher linear algebra\label{app: higher linear alg}}
Here we review some useful results about  (braided)fusion higher categories and their module categories. Most results are due to~\cite{D_coppet_2023_Morita,D_coppet_2023_alg,D_coppet_2023_gauging,Kong_2022_cql1,Kong_2024_cql2}. They can all be viewed as special cases of the higher condensation theory~\cite{kong2025highercondensationtheory}.
\subsection{Morita theory of fusion 2-categories}
\subsubsection{Condensable $E_1$-algebras in fusion 2-categories}
\begin{definition}
    Let $\e{C}$ be a fusion 2-category, an algebra in $\e{C}$ is an object $\cc{A}$ together with a unit $i: I\to \cc{A}$, a product $\mu: \cc{A}\otimes \cc{A}\to \cc{A} $ together certain coherence 2-isomorphisms ensuring the algebra is unital and associative.
\end{definition}
The class of algebras relevant for physics is more restrictive.
\begin{definition}
    An algebra $\cc{A}$ in $\e{C}$ is called rigid if the product $\mu$ admits a right adjoint $\mu^*: \cc{A}\to\cc{A}\otimes \cc{A}$ as an $\cc{A}-\cc{A}$ bimodule 1-morphism.
\end{definition}
\begin{definition}
    A separable algebra in $\e{C}$ is a rigid algebra $\cc{A}$ together with a section of the counit $\mu\circ \mu^*\To\id_{\cc{A}}$ as an $\cc{A}-\cc{A}$ bimodule 2-morphism.
\end{definition}
\begin{definition}
    An algebra $\cc{A}$ in $\e{C}$ is called connected if the unit $i: I\to \cc{A}$ is a simple 1-morphism.
\end{definition}

Separable algebras are used to define Morita equivalence of fusion 2-categories. 

\begin{theorem}
    Let $\cc{A}$ be a separable algebra in $\e{C}$, then $\Bimod_{\cc{A}-\cc{A}}(\e{C})$, the 2-category of $\cc{A}-\cc{A}$ bimodules in $\e{C}$, is a multi-fusion 2-category. If $\cc{A}$ is also connected, then $\Bimod_{\cc{A}-\cc{A}}(\e{C})$ is a fusion 2-category.
\end{theorem}
This leads to the following physical definition.
\begin{definition}
    A condensable $E_1$-algebra in a fusion 2-category $\e{C}$ is a connected separable algebra.  We call $\Bimod_{\cc{A}-\cc{A}}(\e{C})$ the dual fusion 2-category obtained by condensing the algebra $\cc{A}$.
\end{definition}
\subsubsection{Module 2-categories over fusion 2-categories}
\begin{definition}
    Let $\e{C}$ be a fusion 2-category, a left module 2-category is a 2-category $\e{M}$ together with a monoidal 2-functor $\e{C}\to\edo(\e{M})$, i.e. a  coherent left action of $\e{C}$ on $\e{M}$.
\end{definition}
\begin{lemma}
    Let $\e{M}$ be a finite semi-simple module 2-category over $\e{C}$, then there exists a rigid algebra $\cc{A}$ in $\e{C}$, such that $\e{M}\simeq \Mod_{\cc{A}}(\e{C})$, the 2-category of right $\cc{A}$ modules in $\e{C}$.
\end{lemma}
\begin{definition}
    We call a finite semi-simple module 2-category $\e{M}$ separable if the algebra $\cc{A}$ is separable. This does not depend on the choice of $\cc{A}$.
\end{definition}
Denote by $\e{C}_{\e{M}}^*:=\edo_{\e{C}}(\e{M})$ the monoidal 2-category of left $\e{C}$-module 2-functors from $\e{M}$ to $\e{M}$.
\begin{theorem}
    Let $\e{M}\simeq \Mod_{\cc{A}}(\e{C})$ be a separable left module 2-category over $\e{C}$. Then,
    \begin{align}
       \e{C}_{\e{M}}^*\simeq \Bimod_{\cc{A}-\cc{A}}(\e{C})^{op}.
    \end{align}
    as multi-fusion 2-categories.
\end{theorem}
We will say $\e{C}_{\e{M}}^*$ is the dual (multi-)fusion 2-category with respect to $\e{M}$. 

\begin{definition}
    We say two fusion 2-categories $\e{C},\e{D}$ are Morita equivalent if one is the dual fusion 2-category of the other in the above sense.
\end{definition}
\subsection{Relative tensor product of module categories}
The concept of relative tensor product is the mathematical foundation of the \symto sandwich construction. 
\begin{definition}
    Let $\e{M},\e{N}$ be separable right/left module 2-categories over $\e{C}$. Their relative tensor product is
    \begin{align}
\e{M}\boxtimes_{\e{C}}\e{N}=\mathbf{Func}_{\e{C}}(\e{M}^{op},\e{N}).\label{eq: app_relative_1}
    \end{align}
\end{definition}
This is not the usual definition of relative tensor product. Usually one defines relative tensor product as an object with a balanced map from $\e{M}\boxtimes\e{N}$ that is universal among such objects. Then one can prove that if the relative tensor product exists, it is given by Eq.\ref{eq: app_relative_1}. For our purpose Eq.\ref{eq: app_relative_1} suffices. 
\begin{lemma}
    Let $\e{M}=\Mod_{\cc{A}}(\e{C}),~ \e{N}=\Mod_{\cc{B}}(\e{C})$ be separable left $\e{C}$-module 2-categories, then 
    \begin{align}
        \e{M}^{op}\boxtimes_{\e{C}} \e{N}=\Bimod_{\cc{A}-\cc{B}}(\e{C}).\label{eq: app_relative}
    \end{align}
\end{lemma}
\hide{
\subsection{Delooping}
\begin{definition}
    Let $\e{C}$ be a multi-fusion $n$-category, we denote by $\Sigma\e{C}:=\Mod_{\e{C}}((n+1)\vc)$ the finite semi-simple $n+1$-category of separable left $\e{C}$-module $n$-categories, $\e{C}$-module $n$-functors, $\e{C}$-module $n$-natural transformations, etc.
\end{definition}
\begin{lemma}
    Let $\e{B}$ be a braided multi-fusion $n$-category, then every left $\e{B}$-module 2-category is also a right $\e{B}$-module 2-category. Then relative tensor product of module categories makes $\Sigma\e{B}$ a multi-fusion $(n+1)$-category.
\end{lemma}
}
\subsection{Centers}
\begin{definition}
    Let $\e{M}$ be a finite semi-simple $n$-category, its $E_0$-center is 
    \begin{align}
        \Z_0(\e{M}):=\edo(\e{M}),
    \end{align}
    which is a multi-fusion $n$-category.
\end{definition}
\begin{definition}
    Let $\e{C}$ be a multi-fusion $n$-category, its $E_1$-center is 
    \begin{align}
        \Z_1(\e{C}):=\edo_{\e{C}-\e{C}}(\e{C}),
    \end{align}
    which is a braided multi-fusion $n$-category.
\end{definition}
The $E_1$-center is Morita invariant:
\begin{theorem}
    If $\e{C},\e{D}$ are two Morita equivalent fusion $n$-categories, then 
    \begin{align}
        \Z_1(\e{C})\simeq \Z_1(\e{D})
    \end{align}
    as braied fusion $n$-categories.
\end{theorem}
\begin{lemma}
    Let $\e{C}$ be a fusion $n$-category, then,
    \begin{align}
        \Sigma\Z_1(\e{C})\simeq \Z_0(\Sigma\e{C}).
    \end{align}
\end{lemma}
\subsection{$E_1$-fusion module 2-category over braided fusion 2-category}
\begin{definition}
    Let $\e{B}$ be a braided fusion 2-category, a $E_1$-fusion left module 2-category is a fusion 2-category $\e{D}$ together with a braided 2-functor $\e{F}: \e{B}\to \Z_1(\e{D})$.
\end{definition}
\begin{remark}
    Equivalently, we have a monoidal 3-functor $\Sigma\e{F}: \Sigma\e{B}\to \Sigma\Z_1(\e{D})\simeq \Z_0(\Sigma\e{D})$. Therefore a $E_1$-fusion module 2-category $\e{D}$ over $\e{B}$ is the same as a usual module 3-category $\Sigma\e{D}$ over $\Sigma\e{B}$.
\end{remark}
\begin{definition}
    Let $\e{D},\e{E}$ be right/left $E_1$-fusion module 2-category over $\e{B}$, their relative tensor product is 
    \begin{align}
\e{D}\boxtimes_{\e{B}}\e{E}:=\Omega(\Sigma\e{D}\boxtimes_{\Sigma\e{B}}\Sigma\e{E}),
    \end{align}
    which is a multi-fusion 2-category.
\end{definition}
\begin{theorem}
    Let $\e{C}$ be a fusion 2-category, and $\e{M},\e{N}$ be separable right/left  module 2-categories over $\e{C}$, then we have $\Z_1(\e{C}_{\e{M}}^*)\simeq \Z_1(\e{C})\simeq \Z_1(\e{C}_{\e{N}}^*)$, which means the dual fusion 2-categories $\e{C}_{\e{M}}^*,~\e{C}_{\e{N}}^*$ are right/left $E_1$-fusion module 2-categories over $\Z_1(\e{C})$. Then we have 
    \begin{align}
        \e{C}_{\e{M}}^*\boxtimes_{\Z_1(\e{C})}\e{C}_{\e{N}}^*\simeq \Z_0(\e{M}\boxtimes_{\e{C}}\e{N}).\label{eq: app_relative_E1}
    \end{align}
\end{theorem}
\section{Condensable $E_2$-algebras in braided fusion 2-categories\label{app: condensation}}
Here we review the definition of condensable $E_2$-algebras and their (local)modules. 

\begin{definition}\label{def: E_2}
    Let $\e{C}$ be a braided fusion 2-category, an $E_2$-algebra in $\e{C}$ is an object $\e{A}$ together with a 1-morphism called the product: $\mu: \e{A}\otimes \e{A}\to \e{A}$, a 1-morphism called the unit: $i: \bbm{1}\to \e{A}$, together with invertible 2-morphisms in the following diagram:
    \begin{itemize}
        \item The left and right unitors $l,r$:
        \begin{center}
           \begin{tikzcd}
      & \e{A}\otimes \e{A} \arrow[rd,"1"]&\\
       \e{A}\arrow[ru,"i1"] \arrow[rr,equal,""{name=U,above}]& & \e{A} \arrow[Rightarrow, to=1-2, from=U,"l"]
       \end{tikzcd},~  \begin{tikzcd}
      & \e{A}\otimes \e{A} \arrow[rd,"1"]&\\
       \e{A}\arrow[ru,"1i"] \arrow[rr,equal,""{name=U,above}]& & \e{A} \arrow[Rightarrow, from=1-2, to=U,"r"]
       \end{tikzcd}
        \end{center}
       
        \item An associator $a$:
        \begin{center}
            \begin{tikzcd}
                (\e{A}\otimes\e{A})\otimes \e{A} \arrow[rr,equal]\arrow[d,"\mu 1"] && \e{A}\otimes(\e{A}\otimes\e{A})\arrow[d,"1 \mu"]\\
                \e{A}\otimes \e{A}\arrow[dr,"\mu"]&& \e{A}\otimes\e{A}\arrow[dl,"\mu"]\\
                & \e{A}&
                \arrow[from=2-1,to=2-3,"a",Rightarrow,shorten=1.5cm]
            \end{tikzcd}
        \end{center}
        \item A braiding $\beta$:
        \begin{center}
            \begin{tikzcd}
                \e{A}\otimes \e{A}\arrow[rr,"c_{\e{A},\e{A}}",""{name=U, below}] \arrow[dr, "\mu"]&&\e{A}\otimes \e{A}\arrow[dl,"\mu"]\\
                &\e{A}\arrow[from=U,Rightarrow,shorten=1mm,"\beta"]&
                
            \end{tikzcd}
        \end{center}
    \end{itemize}
   These 2-morphisms themselves need to satisfy a set of coherence conditions, which can be found in~\cite{D_coppet_2024}.
\end{definition}

An $E_2$-algebra will play the role of the new vacuum in the condensed phase, therefore we need impose a stability condition of the vacuum as explained in~\cite{kong2025highercondensationtheory}.

\begin{definition}\label{def: c_E_2}
    A condensable $E_2$-algebra in $\e{C}$ is a connected separable $E_2$-algebra. The collection of condensable $E_2$-algebras in $\e{C}$ will be denoted as $\alg^c_{E_2}(\e{C})$.
\end{definition}
Here connected means the unit $i$ is a simple 1-morphism, and separable means the product $\mu$ admits a section(a co-product) as a map between $\e{A}-\e{A}$ bimodules. These details can be found in~\cite{D_coppet_2023_alg}.

\begin{definition}
  Let $\e{A}\in\alg^c_{E_2}(\e{C})$.  A right module over $\e{A}$ is an object $\e{M}\in\e{C}$ together with a 1-morphism $\nu: \e{M}\otimes \e{A}\to \e{M}$, and two invertible 2-morphisms
  \begin{itemize}
      \item A unitor \begin{center}
           \begin{tikzcd}
      & \e{M}\otimes \e{A} \arrow[rd,"1"]&\\
       \e{M}\arrow[ru,"1i"] \arrow[rr,equal,""{name=U,above}]& & \e{M} \arrow[Rightarrow, from=1-2, to=U,""]
       \end{tikzcd}
      \end{center}
      \item An associator  \begin{center}
            \begin{tikzcd}
                (\e{M}\otimes\e{A})\otimes \e{A} \arrow[rr,equal]\arrow[d,"\nu 1"] && \e{M}\otimes(\e{A}\otimes\e{A})\arrow[d,"1 \mu"]\\
                \e{M}\otimes \e{A}\arrow[dr,"\nu"]&& \e{M}\otimes\e{A}\arrow[dl,"\nu"]\\
                & \e{M}&
                \arrow[from=2-1,to=2-3,"",Rightarrow,shorten=1.5cm]
            \end{tikzcd}
        \end{center}
  \end{itemize}
  These 2-morphisms need to satisfy certain coherence conditions, which can be found in~\cite{D_coppet_2023_Morita}. 

\end{definition}
  
  By the results in~\cite{D_coppet_2024}, the collection of all separable right modules over $\e{A}$,denoted as $\Mod_{\e{A}}(\e{C})$, has a natural fusion 2-category structure.
\begin{definition}
      Let $\e{A}\in\alg^c_{E_2}(\e{C})$.  A local module over $\e{A}$ is a right module $M$ together with an invertible 2-morphism in the following diagram called a holonomy
\begin{center}
            \begin{tikzcd}
                \e{M}\otimes \e{A}\arrow[rr,"c_{\e{A},\e{M}}\circ c_{\e{M},\e{A}}",""{name=U, below}] \arrow[dr, "\nu"]&&\e{M}\otimes \e{A}\arrow[dl,"\nu"]\\
                &\e{M}\arrow[from=U,Rightarrow,shorten=1mm,"\hbar"]&
                
            \end{tikzcd}
        \end{center}
      The holonomy needs to satisfy certain conditions, see~\cite{D_coppet_2024}.
\end{definition}
The collection of all local modules over $\e{A}$, denoted as $\Mod^0_{\e{A}}(\e{C})$, has a natural braided fusion 2-category structure.
\section{Extension theory of (braided)fusion categories\label{app: extension theory}}
Here we review the theory of extending (braided)fusion categories by finite groups developed in~\cite{etingof2009fusioncategorieshomotopytheory}.

\begin{definition}\it
The Brauer-Picard 3-group $\underline{\underline{\mathbf{BrPic}}}(\cc{C})$ of a fusion category $\mathcal{C}$ is a 3-category with one object $\mathcal{C}$ and whose

\begin{itemize}
    \item 1-morphisms are invertible $\e{C}-\e{C}$ bimodule categories,
    \item 2-morphisms are equivalences of such bimodule categories,
    \item 3-morphisms are isomorphisms of such equivalences.
\end{itemize}
    
\end{definition}

The homotopy groups of $B\underline{\underline{\mathbf{BrPic}}}(\cc{C})$ are
\begin{align}
    \pi_1=\mathbf{BrPic}(\cc{C}),~\pi_2=\mathbf{Inv}(\Z_1(\cc{C})),~\pi_3=\bC^\times,~\pi_i=0 ~\text{for}~i\ge 4
\end{align}
where $\mathbf{Inv}(\Z_1(\cc{C}))$ stands for the group of invertible objects in $\Z_1(\cc{C})$.

An extension of a fusion category $\mathcal{C}$ by a finite group $G$ is a $G$-graded fusion category whose trivial component  is  $\mathcal{C}$. There is a natural notion of equivalence of $G$-extensions.
\begin{theorem}
    \textit{(a) Equivalence classes of $G$-extensions of $\mathcal{C}$ are given by 3-group homomorphisms from $G$ to} $\underline{\underline{\mathbf{BrPic}}}(\cc{C})$.

\textit{(b) Such homomorphisms (and hence, $G$-extensions) are parameterized by triples $(c, M, \alpha)$, where $c$ is a group homomorphism $c: G \to \underline{\mathbf{BrPic}}(\cc{C})$, $M$ belongs to a certain $H^2(G, \pi_2\underline{\mathbf{BrPic}}(\cc{C}))$-torsor $T_c^2$, and $\alpha$ belongs to a certain $H^3(G, \pi_1\underline{\underline{\mathbf{BrPic}}}(\cc{C}))$-torsor $T_{c,M}^3$.}

\textit{(c) Certain obstruction classes $\cc{O}_3(c) \in H^3(G, \pi_2\underline{\underline{\mathbf{BrPic}}}(\cc{C}))$ and $\cc{O}_4(c,M) \in H^4(G, \pi_3\underline{\underline{\mathbf{BrPic}}}(\cc{C}))$ must vanish for $(c, M, \alpha)$ to determine an extension.}
\end{theorem}

\begin{definition}\it
     Let $\cc{B}$ be a braided fusion category, $\underline{\Aut_{br}}(\cc{B})$ is the 2-group of braided auto-equivalences of $\cc{B}$ and natural isomorphisms. 
\end{definition}
\begin{proposition}\it
  There is a canonical isomorphism of 2-groups
    \begin{align}
       \underline{\mathbf{BrPic}}(\cc{C})\simeq \underline{\Aut_{br}}(\Z_1(\cc{C}))
    \end{align}
\end{proposition}

\begin{definition}\it
  Let $\cc{B}$ be a braided fusion category, then every left module category of $\cc{B}$ is  automatically a $\cc{B}$-bimodule category. The Picard 3-group $\underline{\underline{\mathbf{Pic}}}(\cc{B})$ of a braided fusion category $\cc{B}$ is the 3-category with one object and whose 1-morphisms are invertible left $\cc{B}$-module categories, 2-morphisms are equivalences of module categories, and 3-morphisms are isomorphisms of such equivalences.
\end{definition}
The homotopy groups of $\underline{\underline{\mathbf{Pic}}}(\cc{B})$ are 
\begin{align}
    \pi_1\underline{\underline{\mathbf{Pic}}}(\cc{B})=\mathbf{Pic}(\cc{B}),~\pi_2\underline{\underline{\mathbf{Pic}}}(\cc{B})=\mathbf{Inv}(\cc{B}),~ \pi_3\underline{\underline{\mathbf{Pic}}}(\cc{B})=\bC^\times, \pi_i\underline{\underline{\mathbf{Pic}}}(\cc{B})=0~\text{for}~i\ge4.
\end{align}

A $G$-crossed extension of $\cc{B}$ is a $G$-crossed braided fusion whose trivial component is $\cc{B}$.
\begin{theorem}\it
   Let $\cc{B}$ be a braided fusion category.  Equivalence classes of $G$-crossed extensions of $\cc{B}$ are in bijection with morphisms of 3-groups $G\to \underline{\underline{\mathbf{Pic}}}(\cc{B})$.
\end{theorem}

\end{document}